\documentclass[11pt]{article}
\pdfoutput=1

\usepackage[usenames,dvipsnames,svgnames,table]{xcolor} 
\usepackage[obeyspaces,hyphens,spaces]{url}
\usepackage{jcapmod}
\usepackage[normalem]{ulem}

\usepackage{shorthand}
\usepackage{mathtools}
\usepackage{booktabs}
\usepackage[english]{babel}
\usepackage{amsmath,amssymb,amsbsy,amstext, amsthm, simplewick, amsfonts,braket}
\usepackage{hyperref}
\usepackage{graphicx}
\usepackage[small]{caption}
\usepackage{slashed}
\usepackage[separate-uncertainty=true,multi-part-units=single]{siunitx}
\usepackage{upgreek}
\usepackage{framed}
\usepackage{wrapfig}
\usepackage{multirow}
\usepackage{bbm}
\usepackage{enumerate}

\usepackage[utf8x]{inputenc}
\usepackage{selinput}
\usepackage{bm}
\usepackage{float}
\usepackage{geometry}
\usepackage{yfonts}
\usepackage{caption}
\usepackage{subcaption}
\usepackage{sidecap}
\usepackage{longtable}
\usepackage{anyfontsize}
\usepackage{dsfont}
\usepackage{tikz}
\usepackage{relsize}
\usepackage{tcolorbox}

\usepackage{tikz}
\usetikzlibrary{calc}
\newcommand*\circled[1]{\tikz[baseline=(char.base)]{
    \node[shape=circle, draw, inner sep=1pt, 
        minimum height={\f@size*1.6},] (char) {\vphantom{WAH1g}#1};}}
\makeatother

\newcommand{\minus}{{\scalebox {0.75}[1.0]{$-$}}}

\usepackage{colortbl}
\definecolor{lightgreen}{cmyk}{0.2, 0, 0.2, 0.2}
\definecolor{lightgray2}{cmyk}{0.1,0.1,0,0.1}
\definecolor{Red2}{RGB}{214, 39, 40}
\definecolor{Blue2}{RGB} {31, 119, 180}
\definecolor{Orange2}{RGB}{255, 127, 14}
\definecolor{Green2}{RGB}{44, 160, 44}
\definecolor{greyish2}{rgb}{.96,.96,.96}

\makeatletter
\newlength{\apb@width}
\newcommand{\autoparbox}[2][c]{\settowidth{\apb@width}{#2}\parbox[#1]{\apb@width}{#2}}

\makeatother


\allowdisplaybreaks[1]
\newcommand{\ped}[1]{\textormath{\textsubscript{#1}}{_{\mathrm{#1}}}}

\newcommand{\sped}[1]{_{\textsc{#1}}}

\DeclarePairedDelimiter{\abs}{\lvert}{\rvert}

\renewcommand{\vec}[1]{\boldsymbol{\mathbf{#1}}}

\newcommand{\dd}{\mathop{\mathrm{d}\!}{}}

\DeclareMathOperator\sgn{sgn}

\renewcommand{\Im}{\operatorname{Im}}
\renewcommand{\Re}{\operatorname{Re}}

\definecolor{pyBlue}{RGB}{31, 119, 180}
\definecolor{pyRed}{RGB}{214, 39, 40}
\definecolor{pyGreen}{RGB}{44, 160, 44}
\definecolor{pyBlue2}{RGB}{0, 111, 237}
\definecolor{pyRed2}{RGB}{224, 52, 36}
\definecolor{Mathematica1}{rgb}{0.368417, 0.506779, 0.709798}
\definecolor{Mathematica2}{rgb}{0.880722, 0.611041, 0.142051}
\definecolor{Mathematica3}{rgb}{0.560181, 0.691569, 0.194885}
\definecolor{Mathematica4}{rgb}{0.922526, 0.385626, 0.209179}

\def\beq{\begin{equation}}
\def\eeq{\end{equation}}

\begin{document}

\pagenumbering{roman}
\begin{titlepage}
\baselineskip=15.5pt \thispagestyle{empty}

\phantom{h}
\vspace{1cm}
\begin{center}
{\fontsize{22}{0}\selectfont  \bfseries  Spin-Orbit Alignment of\\\vskip 12pt Gravitational Atoms in Binaries}\\
\end{center}

\vspace{0.5cm}
\begin{center}
{\fontsize{12}{18}\selectfont Giovanni Maria Tomaselli,$^{1}$ Roman R.~Rafikov$^{2}$}

\end{center}

\vspace{-0.1cm}
\begin{center}
\vskip 6pt
\textit{$^1$ School of Natural Sciences, Institute for Advanced Study,\\
1 Einstein Drive, Princeton, NJ 08540, USA}

\vskip 6pt
\textit{$^2$ Department of Applied Mathematics and Theoretical Physics, University of Cambridge,\\
Wilberforce Road, Cambridge CB3 0WA, UK}
\end{center}

\vspace{1.2cm}
\hrule \vspace{0.3cm}
\noindent {\bf Abstract}\\[0.1cm]
Ultralight bosons can form superradiant clouds around rotating black holes (BHs). When the BH is a part of a binary, its companion's tidal field can mix the superradiant state with decaying ones, and deplete the cloud even when the binary separation is away from resonances. We consider orbits inclined with respect to the BH spin, and study the full vectorial evolution of the BH spin, the cloud, and the orbit under the static (i.e., orbit-averaged) part of the tidal potential, finding two distinct regimes. At large separations, the cloud remains locked to the BH spin direction and the system drifts along the superradiant threshold, with two possible outcomes. If the BH spin starts close enough to the orbital axis or its opposite, it aligns or anti-aligns with it; otherwise, the cloud is exhausted first, leaving the BH spin misaligned. At small separations, the cloud orients itself along the orbital axis rather than the BH spin, and the same two outcomes apply. In this case, however, anti-alignment can be unstable, with a violent ``orbit flip'' turning it into alignment. We show that the BH spin-orbit alignment can be faster than the gravitational-wave-driven inspiral for gravitational couplings $\alpha\lesssim0.2$. These results have wide-ranging implications, such as opening up new possibilities to infer the existence of ultralight bosons from the spin tilts of black hole binaries.
\vskip10pt
\hrule
\vskip10pt

\end{titlepage}

\thispagestyle{empty}
\setcounter{page}{2}

\tableofcontents

\newpage
\pagenumbering{arabic}
\setcounter{page}{1}

\clearpage

\section{Introduction}
\label{sec:intro}

The spins of black holes (BHs) carry information about their formation and about fundamental physics. If an ultralight bosonic field exists in nature, spinning BHs are unstable against its perturbations and can transfer a significant fraction of their energy and angular momentum to a cloud of bosons, a process known as \emph{superradiance} \cite{ZelDovich1971,ZelDovich1972,Starobinsky:1973aij,Brito:2015oca}. Such bosons are predicted as a possible solution to the strong CP problem (i.e., the QCD axion \cite{Weinberg:1977ma,Wilczek:1977pj,Peccei:1977hh}), can arise from string compactifications \cite{Arvanitaki:2009fg,Svrcek:2006yi}, and have been proposed as dark matter candidates \cite{Bergstrom:2009ib,Marsh:2015xka,Hui:2016ltb,Ferreira:2020fam}. Superradiance is efficient when the boson's Compton wavelength is comparable to the BH's size, i.e., when the gravitational fine structure constant $\alpha\equiv G\mu M/(\hbar c)$ is of order $\mathcal O(0.1)$, with $\mu$ and $M$ the boson and BH masses. The resulting BH-cloud system is often called a ``gravitational atom,'' as its spectrum closely resembles that of the hydrogen atom \cite{PhysRevD.22.2323,Arvanitaki:2010sy,Baumann:2019eav}. An unstable growing quasi-bound state spins the BH down to its \emph{superradiant threshold}, where the system achieves a long-lived metastable equilibrium. If the BH accretes matter, the gravitational atom then evolves by remaining close to the threshold, moving along it towards larger masses \cite{Brito:2014wla}, while the cloud grows or decays to keep the instability saturated, a behavior known as \emph{threshold drift} \cite{Hui:2022sri}.

\vskip 4pt
When a gravitational atom is part of a binary system, a very rich phenomenology is known to arise \cite{Baumann:2018vus,Zhang:2018kib,Berti:2019wnn,Zhang:2019eid,Baumann:2019ztm,Wong:2020qom,Ikeda:2020xvt,Choudhary:2020pxy,Takahashi:2021eso,Cole:2022yzw,Cao:2023fyv,Duque:2023seg,Guo:2024iye,DellaMonica:2025zby,Guo:2025ckp,Roy:2025qaa}. At specific orbital frequencies, the companion resonantly drives transitions between states of the cloud, known as hyperfine, fine, or Bohr resonances depending on the energy splitting \cite{Baumann:2018vus,Zhang:2018kib,Berti:2019wnn,Baumann:2019ztm,Ding:2020bnl,Takahashi:2021eso,Tong:2021whq,Takahashi:2021yhy,Takahashi:2023flk,Boskovic:2024fga,Tomaselli:2024bdd,Tomaselli:2024dbw,Takahashi:2024fyq,Kyriazis:2025fis,Peng:2025zca,Li:2025qyu,Tomaselli:2025jfo,Boskovic:2025ixx,Su:2026agc,Liang:2026njr,Li:2026wds,Boskovic:2026eth}. A companion close enough to the cloud can also unbind part of it, which is referred to as ionization or dynamical friction \cite{Zhang:2019eid,Takahashi:2021yhy,Baumann:2021fkf,Baumann:2022pkl,Tomaselli:2023ysb,Cao:2024wby,Li:2025qyu,Ding:2025nxe,Cheng:2026dsq}. Accretion onto the companion \cite{Baumann:2021fkf,Guo:2023lbv,Yang:2025gmy} and the cloud's self-gravity \cite{Ferreira:2017pth,Hannuksela:2018izj,Su:2021dwz,Cao:2024wby,Kim:2025wwj} have also been shown to be relevant. Some of these phenomena have recently been studied with a fully relativistic formalism \cite{Brito:2023pyl,Khalvati:2024tzz,Dyson:2025dlj,Li:2025ffh,Cannizzaro:2025vpb,Xu:2026cky,Xu:2026aic,Li:2026wds,Guo:2026qfb,Xu:2026qrg,Li:2026foy}.

\vskip 4pt
In classical celestial mechanics, a companion's perturbing potential can be divided into three types of terms \cite{MurrayDermott1999,2023dyps.book.....T}. \emph{Secular} terms do not depend on the companion's orbital phase and are obtained by averaging its gravitational potential over the orbit. Time-dependent terms are called \emph{resonant} if their frequency nearly matches a natural frequency of the perturbed system, and \emph{short-period} otherwise. To leading order, short-period terms average out and are routinely neglected \cite{MurrayDermott1999,Silsbee:2013yaa}. Resonant terms act only near discrete orbital separations, such as those of the resonances above. Secular terms, on the other hand, are global: they act at every separation, even away from resonances, provided the perturbed system evolves slowly compared with the orbital period. Classic examples are the (Laplace-Lagrange) secular evolution of planetary orbits \cite{MurrayDermott1999},  the von Zeipel--Lidov--Kozai oscillations of hierarchical triples \cite{1910AN....183..345V,Kozai:1962zz,Lidov:1962wjn} and the evolution of binaries in the tidal field of a star cluster \cite{2019MNRAS.488.5489H}.

\vskip 4pt
For the gravitational atom, however, secular terms have so far received comparatively little attention. The static tidal deformation of the cloud has been quantified through its Love numbers \cite{Cardoso:2020hca,DeLuca:2021ite,Arana:2024kaz,Boskovic:2026eth}. The companion's tide is also known to mix the superradiant state with decaying ones, depleting the cloud even away from resonances \cite{Arvanitaki:2009fg,Arvanitaki:2010sy,Arvanitaki:2014wva,Baumann:2018vus,Tong:2022bbl,Takahashi:2023flk,Takahashi:2024fyq,Zhu:2024bqs,DellaMonica:2025zby,Tomaselli:2025jfo}. On eccentric and inclined orbits (relative to BH spin), this depletion was found in Ref.~\cite{Fan:2023jjj} to tilt the orbital plane towards the BH's equator, with the BH spin held fixed. Non-resonant mixing also enters the evolution equations for the spin-orbit misalignment derived in \cite{Boskovic:2025ixx}, but this study assumed that the spins of the cloud and of the BH remain parallel, which we show not to be the case in general. None of these works treats the orientations of the BH spin and of the cloud as independent dynamical variables.

\vskip 4pt
In this paper, we determine the dynamics and long-term evolution of a gravitational atom under the static part of the tidal perturbation of a companion, starting from a given initial state. We follow the full vectorial evolution of the BH spin, the cloud, and the orbit, neglecting the other phenomena mentioned above such as resonances and ionization. When the BH spin and orbital angular momentum are misaligned, the tide mixes the superradiant state with decaying states of the same multiplet, and their horizon absorption acts as internal dissipation. As the cloud depletes due to this process and transfers mass and angular momentum to the BH, conservation of the total angular momentum forces a change in the obliquity, the angle between the BH spin and the orbital angular momentum.

\vskip 4pt
This mechanism bears similarity to tidal evolution in close binary stars, where the tide raised by the companion is dissipated by turbulent viscosity or radiative damping \cite{Zahn:1977mi}. Tidal friction circularizes the orbit and synchronizes the stellar spins, aligning them with the orbital angular momentum \cite{1981A&A....99..126H,Zahn:2008fk}: this is the only configuration in which a binary can be in tidal equilibrium \cite{1980A&A....92..167H}. The closest analog to our setup is the component of the tide that is static in the inertial frame, whose dissipation damps the stellar obliquity without orbital decay, and can drive retrograde systems towards anti-alignment \cite{Lai:2011vr}. Unlike the full stellar tidal interaction, the static tide we consider in this paper conserves the orbit's semi-major axis, eccentricity, and angular momentum magnitude, so only its orientation can change. In the gravitational atom, the tide is dissipated by horizon absorption. The cloud, however, is a finite reservoir of mass and angular momentum, and its exhaustion before alignment has no direct counterpart in stellar binaries.

\vskip 4pt
Tidal alignment in stellar binaries is also important for predicting the spins of merging BH binaries. In the isolated formation channel, mass transfer and tides are expected to align the spins of the BH progenitors with the orbit, while supernova kicks can tilt the orbital plane \cite{Kalogera:1999tq,Gerosa:2013laa,Gerosa:2018wbw}. Binaries assembled dynamically should instead have isotropic spin orientations, so the distribution of spin tilts is a diagnostic of the formation channel \cite{Rodriguez:2016vmx}. This distribution is now being constrained with the BH binaries observed by the LIGO-Virgo-KAGRA collaboration \cite{KAGRA:2021duu,LIGOScientific:2025pvj}. With this paper, we provide a new mechanism that influences this distribution, and can create a preference for (anti-)aligned spins even in binaries assembled dynamically, opening up a new possibility to infer the existence of ultralight bosons from populations of binary BHs. Boson clouds were already shown to leave an imprint on the eccentricity and obliquity of BH binaries through resonant transitions during the inspiral \cite{Boskovic:2024fga,Tomaselli:2024dbw,Boskovic:2025ixx}. Our mechanism is different because it does not involve any resonance.

\vskip 4pt
To study this process, we start by deriving the evolution equations for the cloud, the BH, and the orbit, including their mutual backreaction and geodetic precession. The dynamics is set by the competition between the tidal perturbation and the Lense--Thirring (or hyperfine) coupling of the cloud to the BH spin. When the latter dominates, which is the case at larger separations, the cloud stays aligned with the BH spin and the two precess together about the total angular momentum, a regime we call \emph{locked co-precession}. The BH then evolves by drifting along its superradiant threshold, and the system of equations can be reduced to a one-dimensional flow, whose final state follows directly from conservation laws. At smaller separations, the companion's tide dominates, and the cloud is aligned with the orbital axis rather than with the BH spin. In this regime of \emph{unlocked alignment}, no superradiant threshold exists, so the cloud depletes on the timescale set by horizon absorption.

\vskip 4pt
In both regimes, we find two mutually exclusive final states: either BH \emph{spin-orbit alignment} (or anti-alignment) with a surviving cloud, or \emph{cloud exhaustion}, generally at an intermediate BH spin obliquity. In the locked regime, the system (anti-)aligns if its initial obliquity is close enough to $0$ (or $\pi$), within a range that widens with the cloud's mass; the cloud is exhausted otherwise. Orbital backreaction favors alignment and, when strong enough, leads to it for any initial obliquity. In the unlocked regime, anti-alignment is unstable when the orbital angular momentum is smaller than the cloud's spin, a situation possible for e.g. extreme mass ratio inspirals: systems approaching it undergo an ``orbit flip'' and end up aligned. Direct numerical integrations of the full evolution equations confirm these predictions. Finally, we compare the alignment timescale with the gravitational-wave (GW) merger time: for comparable masses and a cloud in the $\ket{211}$ state, the former is shorter at some stage of the inspiral if $\alpha\lesssim0.2$.

\paragraph{Outline.} The outline of the paper is as follows. In Sec.~\ref{sec:equations}, we introduce the orbit-averaged tidal perturbation and derive the master equations together with their conservation laws. In Secs.~\ref{sec:locked} and~\ref{sec:unlocked}, we study locked co-precession and unlocked alignment, respectively, and characterize their final states and alignment timescales. In Sec.~\ref{sec:numerics}, we test these results by integrating the master equations numerically. In Sec.~\ref{sec:alignment-merger}, we discuss the competition between alignment and the GW-driven inspiral. We conclude in Sec.~\ref{sec:conclusions}.

\paragraph{Notation and conventions.} We work in natural units, $G=\hbar=c=1$.

\section{Evolution equations}
\label{sec:equations}

In this section we derive the coupled evolution equations of the cloud, the BH, and the orbit. In Secs.~\ref{sec:setup}--\ref{sec:timescales}, we set up the problem, average the tidal perturbation over the orbit, and introduce the basic timescales, whose hierarchy defines the regimes studied in Secs.~\ref{sec:locked} and~\ref{sec:unlocked}. We then write the Schrödinger equation in the non-inertial BH-aligned and orbit-aligned frames in Sec.~\ref{sec:non-inertial}, and compute the backreaction on the BH and on the orbit, including geodetic precession, in Sec.~\ref{sec:BHback}. After identifying the conserved quantities in Sec.~\ref{sec:integrals}, we determine the angular velocities of the frames and the evolution of the obliquity in Sec.~\ref{sec:gauge}, and the azimuth of the BH spin in Sec.~\ref{sec:azimuth}. Finally, we collect the master equations in Sec.~\ref{sec:master-equations}.

\subsection{Setup}
\label{sec:setup}

\begin{figure}
\centering
\includegraphics[scale=1.15]{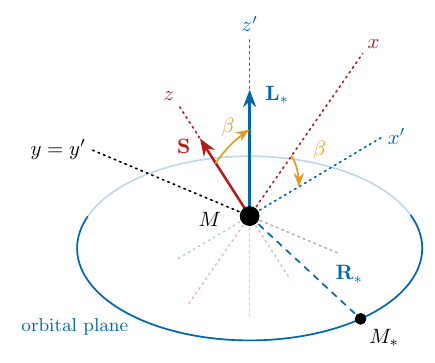}
\caption{Geometry and instantaneous reference frames. The BH of mass $M$ is at a focus of the companion's orbit (blue ellipse). Its spin $\vec S$ defines $\hat z$, while the orbital angular momentum $\vec L_*$ defines $\hat z'$. The two frames share $\hat y=\hat y'$ and are rotated with respect to each other by the obliquity $\beta$.}
\label{fig:frames}
\end{figure}

Consider a BH of mass $M$ and dimensionless spin $\tilde a$, surrounded by a scalar cloud of mass $M\ped{c}$. The BH spin is $\vec S=\tilde aM^2\hat S$, and we denote the cloud's angular momentum by $\vec S\ped{c}$. A companion of mass $M_*$ orbits the gravitational atom with semi-major axis $a$ and eccentricity $e$. We treat the BH and cloud as one binary component and define the mass ratio and orbital angular momentum as
\beq
q\equiv\frac{M_*}{M+M\ped{c}}\,,\qquad
\vec L_*=\sqrt{\frac{q^2(M+M\ped{c})^3a(1-e^2)}{1+q}}\,\hat L_*\,.
\label{eqn:orbital-definitions}
\eeq
The cloud-to-BH mass ratio is $X\equiv M\ped{c}/M$, thus $M_*=q(1+X)M$. We define the \emph{obliquity} $\beta$ as the angle between the BH spin and the orbital angular momentum, $\hat S\cdot\hat L_*=\cos\beta$. We fix a BH-aligned Cartesian frame $xyz$, such that $\hat z=\hat S$ and $\hat L_*$ lies in the $xz$ plane, with a positive $x$ component, see Figure~\ref{fig:frames}. We also define an orbit-aligned frame $x'y'z'$ by rotating these axes through $\beta$ about their common $y$ axis,
\beq
\hat z'=\sin\beta\,\hat x+\cos\beta\,\hat z=\hat L_*\,,\qquad
\hat x'=\cos\beta\,\hat x-\sin\beta\,\hat z\,,\qquad
\hat y'=\hat y=\frac{\hat S\times\hat L_*}{\sin\beta}\,.
\label{eqn:axes}
\eeq

\vskip 4pt
We denote by $\mu$ the mass of the scalar field and by $\alpha\equiv\mu M$ the ``gravitational fine structure constant''. The cloud's characteristic radius is $r\ped{c}\equiv M/\alpha^2$. We consider the case of negligible self-interactions, which can otherwise significantly alter the cloud's evolution and structure \cite{Baryakhtar:2020gao,Takahashi:2024fyq,Witte:2024drg,
Cannizzaro:2026qgf}, and also neglect the cloud's depletion due to self-gravity \cite{Foster:2026nuz,Boskovic:2026boh} and gravitational-wave emission \cite{Yoshino:2013ofa,Brito:2014wla}. In the nonrelativistic limit and in a frame freely falling with the BH, the scalar field's wavefunction $\ket{\psi}$ obeys Schrödinger's equation,
\beq
\frac{\partial}{\partial t}\ket{\psi}=-i(\mathcal H_{\mathrm{ff}}+V_{\mathrm{ff}})\ket{\psi}\,,
\label{eqn:schrodinger}
\eeq
where $\mathcal H_{\mathrm{ff}}$ is the isolated-BH Hamiltonian and $V_{\mathrm{ff}}$ is the gravitational perturbation from the binary companion, both expressed in the freely falling frame.  In the BH spin-aligned frame $xyz$ (which is different from the free-falling frame, see Sec.~\ref{sec:non-inertial}), we denote their representations by $\mathcal H$ and $V$, and the wavefunction by $\ket{c}$. We will derive the transformation of Eq.~\eqref{eqn:schrodinger} between these frames in Sec.~\ref{sec:non-inertial}. We label the eigenstates of $\mathcal H$ by hydrogenic states $\ket{n\ell m}$, with $n>\ell\ge\abs{m}$ and complex frequencies $\omega_{n\ell m}=E_{n\ell m}+i\Gamma_{n\ell m}$. At small $\alpha$, these can be approximated analytically as~\cite{PhysRevD.22.2323,Baumann:2019eav}
\begin{align}
E_{n\ell m}&=\mu\biggl[1-\frac{\alpha^2}{2n^2}-\frac{\alpha^4}{8n^4}
-\frac{(3n-2\ell-1)\alpha^4}{n^4(\ell+1/2)}
+\frac{2\tilde a m\alpha^5}{n^3\ell(\ell+1/2)(\ell+1)}+\mathcal O(\alpha^6)\biggr]\,,
\label{eqn:E_nlm}\\
\Gamma_{n\ell m}&\simeq\frac{2^{4\ell+2}(n+\ell)!}{n^{2\ell+4}(n-\ell-1)!}
\biggl[\frac{\ell!}{(2\ell)!(2\ell+1)!}\biggr]^2
\bigl(1+\sqrt{1-\tilde a^2}\bigr)(m\Omega_+-E_{n\ell m})\alpha^{4\ell+5}g_{\ell m}\,,
\label{eqn:Gamma_nlm}
\end{align}
where $r_+=M(1+\sqrt{1-\tilde a^2})$ is the horizon radius,
$\Omega_+=\tilde a/(2r_+)$ is its angular velocity, and
\beq
g_{\ell m}=\prod_{k=1}^{\ell}\bigl[k^2(1-\tilde a^2)+(\tilde am-2r_+E_{n\ell m})^2\bigr]\,.
\eeq
For arbitrary $\alpha$, the frequencies $\omega_{n\ell m}$ can be determined numerically using Leaver's method~\cite{Leaver:1985ax,Dolan:2007mj,Baumann:2019eav}. In this paper, we rely on \eqref{eqn:E_nlm}--\eqref{eqn:Gamma_nlm} for explicit analytical estimates, and on Leaver's method for numerical integrations of the evolution of the system. Whenever possible, we assume that the \emph{unperturbed} cloud sits in a single pure state $\ket{a}\equiv\ket{n_a\ell_am_a}$, for example one of the fastest-growing states having $\ell_a=m_a$, like $\ket{211}$ or $\ket{322}$.\footnote{In the present section, for notational convenience we will still use the label $\ket{a}$ as a dummy index. From Sec.~\ref{sec:locked} onwards, $\ket{a}$ will be reserved for the most populated state of the cloud.} When the superradiant instability is saturated, such states have purely real energy, $\Gamma_a=0$. Approximating $E_{n\ell m}\approx\mu$ in \eqref{eqn:Gamma_nlm}, this condition can be written as
\beq
\tilde a\big|_{\Gamma_a=0}\approx\frac{4m_a\alpha}{m_a^2+4\alpha^2}\,,
\label{eqn:threshold}
\eeq
to which we will refer as the \emph{superradiant threshold} in the $\alpha$--$\tilde a$ plane. As we show later, however, in some scenarios the perturbation to the cloud can never be considered small, in which case it is useful to consider the wavefunction $\ket{c'}$ in the $x'y'z'$ frame, whose eigenstates we also denote with a primed symbol and azimuthal number, $\ket{a'}\equiv\ket{n_a\ell_am_a'}$. These have their angular momentum quantized about $\hat L_*$, rather than $\hat S$.

\vskip 4pt
We expand the perturbed cloud's wavefunction as $\ket{c}=\sum_ac_a\ket{a}$ in the BH spin-aligned frame, or $\ket{c'}=\sum_{a'}c_{a'}\ket{a'}$ in the orbit-aligned frame. The eigenstates and wavefunction coefficients in the two frames can be related to each other by
\beq
\ket{n\ell m'}=\sum_m d^{(\ell)}_{mm'}(\beta)\ket{n\ell m}\,,\qquad
c_{n\ell m}=\sum_{m'}d^{(\ell)}_{mm'}(\beta)c_{n\ell m'}\,,
\label{eqn:rotated-nlm}
\eeq
where $d^{(\ell)}(\beta)=e^{-i\beta L_y}$ is the real Wigner small $d$-matrix, and $L_j$ denotes the angular momentum operator. The matrix elements of $\vec L$ between two states $\ket{a}$ and $\ket{b}$ in the $xyz$ frame are
\begin{align}
\label{eqn:Lz-matrix-elements}
(L_z)_{ab}&=m_a\delta_{ab}\,,\\
(L_x\pm iL_y)_{ab}&=\sqrt{\ell_a(\ell_a+1)-m_b(m_b\pm1)}\,
\delta_{n_an_b}\delta_{\ell_a\ell_b}\delta_{m_a,m_b\pm1}\,,
\label{eqn:Lxy-matrix-elements}
\end{align}
and analogous expressions hold in the $x'y'z'$ frame, with $m$ replaced by $m'$.

\vskip 4pt
Let $M\ped{c,0}$ be the cloud's mass at an initial time where $\sum_a|c_a|^2=1$. The subsequent cloud's mass and spin in the BH-aligned frame are
\beq
M\ped{c}=M\ped{c,0}\sum_a|c_a|^2\,,\qquad
\vec S\ped{c}=\frac{M\ped{c,0}}\mu\braket{c|\vec L|c}=\frac{M\ped{c,0}}\mu\sum_{a,b}c_a^*\vec L_{ab}c_b\,.
\label{eqn:cloud-observables}
\eeq
A pure BH-aligned state $\ket{a}$ has $\vec S\ped{c}=m_aM\ped{c}\hat S/\mu=(m_a X/\alpha)M^2\hat S$, whereas a pure orbit-aligned state $\ket{a'}$ has $\vec S\ped{c}=m'_aM\ped{c}\hat L_*/\mu=(m'_aX/\alpha)M^2\hat L_*$. 

\subsection{Orbit-averaged tidal perturbation}
\label{sec:tidal-perturbation}

Let $\vec R_*=(R_*,\theta_*,\phi_*)$ be the companion's position relative to the BH, where we used spherical coordinates $(r,\theta,\phi)$ in the BH-aligned frame $xyz$. The companion contributes to the Hamiltonian a potential energy
\beq
V(t,\vec r)=-\frac{M_*\mu}{|\vec r-\vec R_*|}
+\frac{M_*\mu}{R_*^3}\,\vec R_*\cdot\vec r\,,
\label{eqn:V}
\eeq
where the second term is the fictitious force due to the BH's acceleration, also known as the ``indirect potential''. The companion's potential is a periodic function of time, varying over the orbital period $T=2\pi\sqrt{a^3/[(1+q)(M+M\ped{c})]}$. Its oscillating components can induce a wealth of resonant phenomena, driving state transitions at binary separations where eigenenergy splittings match harmonics of the orbital frequency~\cite{Baumann:2018vus,Berti:2019wnn,Baumann:2019ztm,Tomaselli:2024bdd,Tomaselli:2025jfo,Boskovic:2025ixx}.

\vskip 4pt
In this paper, however, we do not focus on that rich phenomenology. Instead, we consider off-resonant dynamics, where the oscillations of the potential average out. We thus only retain the orbital average of the perturbation, which we denote with an overline,
\beq
\bar V(\vec r)\equiv\frac1T\int_0^T V(t,\vec r)\dd t\,,\qquad
\bar V_{ab}\equiv\braket{a|\bar V|b}\,.
\label{eqn:Vbar}
\eeq
We hold the orbital elements and the wavefunction coefficients fixed during this average. This requires their evolution to be slow compared with the orbital period $T$, meaning that the semi-major axis must be smaller than that corresponding to the resonant orbit between the two states considered, i.e., $2\pi/T\gg(E_a-E_b)/(m_a-m_b)$. For states with different $n$ or $\ell$, the resonant orbit is typically located close to the cloud's peak density, where other phenomena like ionization occur, complicating the physical picture. For this reason, in this paper we focus on states within the same $(n,\ell)$ multiplet, having hyperfine splittings given by the last term in \eqref{eqn:E_nlm}, and resonant orbits much wider than the size of the cloud. This is a good approximation for $\ket{a}=\ket{211}$, because the other states connected to it by the quadrupolar tide have $n\ge3$, which suppresses their population, as we will see in Sec.~\ref{sec:adiabatic-elimination}. So we retain $\ket{a}=\ket{211}$ as the main physical case, and require
\beq
a\ll\bigg(\frac{(1+q)(1+X)M}{(E_{211}-E_{210})^2}\bigg)^{1/3}\simeq3.8\times10^3M\,\big[(1+q)(1+X)\big]^{1/3}(1+4\alpha^2)^{2/3}\bigg(\frac{0.2}\alpha\bigg)^{14/3}\,,
\label{eqn:a-hyperfine}
\eeq
where we evaluated the hyperfine splitting from \eqref{eqn:E_nlm} on the superradiant threshold \eqref{eqn:threshold}. For $\ket{a}=\ket{322}$, instead, the coupling to $\ket{300}$ at separations larger than the $\ket{322}$--$\ket{300}$ resonance is known to lead to a quick dissipation of the cloud~\cite{Fan:2023jjj}. We nevertheless retain the possibility of $\ket{a}=\ket{322}$ throughout, as an illustration of the formalism, keeping in mind that the inclusion of $\ket{300}$ would be required to get the correct dynamics.

\vskip 4pt
Since we assume that the companion orbits well outside the cloud, and the cloud's characteristic size is $\braket{a|r^{-1}|a}^{-1}=n_a^2r\ped{c}$, we require $a(1-e)\gg n_a^2r\ped{c}$. Expanding \eqref{eqn:V} in powers of $r/R_*$, the monopole gives an irrelevant global energy shift, the dipole cancels with the indirect potential, and the leading remaining term is the quadrupole,
\beq
V\simeq-\frac{M_*\mu}{2R_*^3}\big(3(\hat R_*\cdot\vec r)^2-r^2\big),
\eeq
whose orbit average can be calculated as
\beq
\overline{\frac{\hat R_{*i}\hat R_{*j}}{R_*^3}}
=\frac{\delta_{ij}-(\hat L_*)_i(\hat L_*)_j}{2a^3(1-e^2)^{3/2}}\implies\bar V(\vec r)\simeq\frac{M_*\mu r^2}{4a^3(1-e^2)^{3/2}}
\bigl[3(\hat r\cdot\hat L_*)^2-1\bigr]\,.
\label{eqn:Vbar-quad}
\eeq
With the conventions of Fig.~\ref{fig:frames}, where $\hat L_*$ lies in the $xz$ plane, we have $\hat r\cdot\hat L_*=\cos\beta\cos\theta+\sin\beta\sin\theta\cos\phi$. Note that at quadrupole order $\bar V(\vec r)$ is independent of the orientation of the orbital apsidal line. Also, it is invariant if $\hat L_*$ is flipped to $-\hat L_*$. Matrix elements of higher multipoles are suppressed by additional powers of $r\ped{c}/(a(1-e))$. In this paper, we restrict to~\eqref{eqn:Vbar-quad}.

\subsection{Timescales}
\label{sec:timescales}

It is useful to introduce the basic timescales entering the problem. Let us start with the characteristic time associated with the tidal perturbation. The potential \eqref{eqn:Vbar-quad} is proportional to $r^2P_2(\hat r\cdot\hat L_*)$, with $P_2$ a Legendre polynomial, that is, to the $m'=0$ component of a rank-2 spherical tensor with respect to the orbit-aligned axis. By the Wigner--Eckart theorem \cite{wigner}, within a given multiplet $(n,\ell)$ the matrix elements of $\bar V$ are proportional to those of the rank-2 tensor built from the angular momentum operator, $3L_{z'}^2-L^2$, with a reduced matrix element that depends on $(n,\ell)$ only. Because $L^2=\ell(\ell+1)$ is constant on the multiplet, $\bar V$ must then be a quadratic polynomial of the form
\beq
\bar V=\mathcal A-\frac{\mathcal C}{T\ped{tidal}}L_{z'}^2\,,
\label{eqn:Vbar-L*}
\eeq
where $\mathcal A$ is an irrelevant multiple of the identity, $\mathcal C$ is a constant that depends on $(n,\ell)$, and we defined the tidal timescale
\beq
\begin{split}
T\ped{tidal}&\equiv\frac{\alpha^3a^3(1-e^2)^{3/2}}{M_*M}
=\frac{\alpha^3a^3(1-e^2)^{3/2}}{q(1+X)M^2}\\
&=\SI{1.25e-5}{yr}\,\frac{(1-e^2)^{3/2}}{q(1+X)}\bigg(\frac{M}{10M_\odot}\bigg)\bigg(\frac{a/M}{10^3}\bigg)^3\bigg(\frac{\alpha}{0.2}\bigg)^3\,.
\end{split}
\label{eqn:T_tidal}
\eeq
Up to $\mathcal O(1)$ factors, $T\ped{tidal}$ can also be understood as the nodal precession timescale of a test-particle orbit of radius $r\ped{c}$ induced by the companion, $T\ped{tidal}\sim(M/M_*)(a/r\ped{c})^3\sqrt{r\ped{c}^3/M}$~\cite{MurrayDermott1999}. Equations \eqref{eqn:Vbar-L*}--\eqref{eqn:T_tidal} can be used to quickly calculate the matrix elements of $\bar V$ between $\ket{a}$ (for which $\ell_a=m_a$) and its neighbors $\ket{b_i}\equiv\ket{n_a\ell_a,\ell_a-i}$. Using $L_{z'}=L_z\cos\beta+L_x\sin\beta$ and the matrix elements \eqref{eqn:Lz-matrix-elements}--\eqref{eqn:Lxy-matrix-elements}, we have
\beq
\bar V_{b_1a}=-\frac{\mathcal C_1}{T\ped{tidal}}\sin\beta\cos\beta\,,\qquad\bar V_{b_2a}=-\frac{\mathcal C_2}{T\ped{tidal}}\sin^2\beta\,,
\label{eqn:Vb1a-Vb2a}
\eeq
with $\mathcal C_1=\mathcal C(2\ell_a-1)\sqrt{\ell_a/2}$ and $\mathcal C_2=\mathcal C\sqrt{(2\ell_a-1)\ell_a}/2$. For $\ket{a}=\ket{211}$ and $\ket{322}$, using hydrogenic radial integrals gives
\beq
\begin{array}{c|ccc}
\ket{a}&\mathcal C_1&\mathcal C_2&\mathcal C\\ \hline
\ket{211}&9/\sqrt2&9/2&9\\
\ket{322}&27&27/\sqrt6&9
\end{array}\,.
\label{eqn:tidal-coefficients}
\eeq
States $\ket{b_i}$ with $i\ge3$ do not directly couple to $\ket{a}$ through $\bar V$.

\vskip 4pt
There is a second important timescale associated with the dynamics of the $(n,\ell)$ multiplet. The spin-dependent energy splitting in \eqref{eqn:E_nlm}, also known as \emph{hyperfine splitting}, describes the Lense--Thirring interaction between the BH and the cloud. This energy splitting can be recovered introducing the Lense--Thirring Hamiltonian
\beq
\mathcal H\sped{lt}=\frac{2\vec S\cdot\vec L}{r^3}\,,\qquad
\braket{n\ell m|\mathcal H\sped{lt}|n\ell m}
=\mu\frac{2\tilde a m\alpha^5}{n^3\ell(\ell+1/2)(\ell+1)}\,.
\label{eqn:hyperfine}
\eeq
This is proportional to $m$, so within a given $(n,\ell)$ multiplet we can write $\mathcal H\sped{lt}=L_z/T\sped{lt}$, where we defined the characteristic Lense--Thirring timescale
\beq
\begin{split}
T\sped{lt}&\equiv\frac{1}{E_{n\ell m}-E_{n\ell,m-1}}
=\frac{Mn^3\ell(\ell+1/2)(\ell+1)}{2\tilde a\alpha^6}\\
&\simeq\begin{cases}
\dfrac{3M(1+4\alpha^2)}{\alpha^7}=\SI{3.66e-7}{yr}\,\bigg(\dfrac{M}{10M_\odot}\bigg)\bigg(\dfrac{0.2}{\alpha}\bigg)^7(1+4\alpha^2)&\ket{211}\,,\\[8pt]
\dfrac{405M(1+\alpha^2)}{4\alpha^7}=\SI{1.23e-5}{yr}\,\bigg(\dfrac{M}{10M_\odot}\bigg)\bigg(\dfrac{0.2}{\alpha}\bigg)^7(1+\alpha^2)&\ket{322}\,.
\end{cases}
\end{split}
\label{eqn:T_lt}
\eeq
The final explicit expressions in \eqref{eqn:T_lt} make use of the superradiant spin-threshold condition given in \eqref{eqn:threshold}. The ratio of the two timescales discussed so far,
\beq
\frac{T\sped{lt}}{T\ped{tidal}}\simeq\frac{q(1+X)}{(1-e^2)^{3/2}}\bigg(\frac{0.2}{\alpha}\bigg)^{10}\bigg(\frac{10^3}{a/M}\bigg)^{3}\times\begin{cases}
0.0293(1+4\alpha^2)&\ket{211}\,,\\[4pt]
0.989(1+\alpha^2)&\ket{322}\,,
\end{cases}
\label{eqn:timescales-ratio}
\eeq
will turn out to be the parameter separating the regimes of Secs.~\ref{sec:locked} and~\ref{sec:unlocked}.

\vskip 4pt
The third basic timescale is that of horizon absorption of the cloud's decaying states. For concreteness, we define it as the decay time of $\ket{b_1}$ when the BH sits on the superradiant threshold of state $\ket{a}$ which is neutral,
\beq
T\ped{abs}\equiv\frac{1}{\abs{\Gamma_{b_1}}}\bigg|_{\Gamma_a=0}\,.
\label{eqn:T_abs}
\eeq
States $\ket{b_i}$ with $i\ge2$ have a decay timescale of comparable order. Evaluating \eqref{eqn:T_abs} on the threshold \eqref{eqn:threshold} and using the analytic approximation \eqref{eqn:Gamma_nlm} yields
\beq
\begin{split}
T\ped{abs}\simeq\begin{cases}
\dfrac{12M(1+4\alpha^2)}{\alpha^{10}}=\SI{1.83e-4}{yr}\,\bigg(\dfrac{M}{10M_\odot}\bigg)\bigg(\dfrac{0.2}{\alpha}\bigg)^{10}(1+4\alpha^2) & \ket{211}\,,\\[8pt]
\dfrac{5\cdot3^{11}M}{2^5\alpha^{14}}\dfrac{(1+\alpha^2)^3}{1-\alpha^2+\alpha^4}=\SI{2.64e2}{yr}\,\bigg(\dfrac{M}{10M_\odot}\bigg)\bigg(\dfrac{0.2}{\alpha}\bigg)^{14}\dfrac{(1+\alpha^2)^3}{1-\alpha^2+\alpha^4} & \ket{322}\,.
\end{cases}
\end{split}
\label{eqn:T-abs-explicit}
\eeq
These expressions are only accurate at small $\alpha$: for instance, an error of less than $10\%$ requires $\alpha\lesssim0.05$. At moderate or large $\alpha$, numerical evaluation through Leaver's method gives more precise results.

\vskip 4pt
These three timescales, $T\ped{tidal}$, $T\sped{lt}$, and $T\ped{abs}$ already hint at the regimes and dynamics we will discuss in Secs.~\ref{sec:locked} and \ref{sec:unlocked}. When $T\sped{lt}\ll T\ped{tidal}$, the companion's perturbation is weak enough that the cloud can remain close to a pure unperturbed eigenstate in the BH-aligned frame. Conversely, if $T\ped{tidal}\ll T\sped{lt}$, the Lense--Thirring precession acts as a weak perturbation to the orbit-aligned eigenstates. In either case, $T\ped{abs}$ determines the exchange of mass and angular momentum between BH and cloud, which ultimately drives the evolution of the system.

\subsection{Non-inertial Schrödinger equation}
\label{sec:non-inertial}

The tidal perturbation mixes states of the cloud, which then exchange mass and angular momentum with the BH. The BH spin and the orbital plane can therefore move, so neither the $xyz$ nor the $x'y'z'$ frame is generally inertial. We thus need to determine how the Schrödinger equation changes when written in either of these frames.

\vskip 4pt
We use as a reference the freely falling frame in which $\ket\psi$ was defined in Sec.~\ref{sec:setup}. Its axes themselves undergo geodetic, or de Sitter, precession relative to a fixed external frame. Treating the whole gravitational atom as one binary component, the leading orbit-averaged angular velocity of this precession is~\cite{Barker:1975ae,Kidder:1995zr}
\beq
\vec\Omega\ped{dS}=\gamma\ped{dS}\vec L_*\,,\qquad
\gamma\ped{dS}\equiv\frac{2+3q/2}{a^3(1-e^2)^{3/2}}\,.
\label{eqn:geodetic}
\eeq
Working in the freely falling frame removes this precession from the local equations of motion. We denote by $\vec\Omega$ and $\vec\Omega'$ the angular velocities of $xyz$ and $x'y'z'$ \emph{relative to the freely falling frame}. Their angular velocities relative to external axes would require the addition of $\vec\Omega\ped{dS}$.

\vskip 4pt
Let $\mathcal R(t)$ be the unitary rotation that maps wavefunctions $\ket{c}$ in the BH-aligned frame into wavefunctions $\ket{\psi}$ in the freely falling one, so that $\ket\psi=\mathcal R\ket c$. The operators introduced in Sec.~\ref{sec:setup} are then related by $\mathcal H_{\mathrm{ff}}=\mathcal R\mathcal H\mathcal R^\dagger$, and $\bar V_{\mathrm{ff}}=\mathcal R\bar V\mathcal R^\dagger$, and the orbit-averaged Schrödinger equation reads
\beq
\frac{\partial}{\partial t}\ket\psi
=-i(\mathcal H_{\mathrm{ff}}+\bar V_{\mathrm{ff}})\ket\psi=-i\mathcal R(\mathcal H+\bar V)\mathcal R^\dagger\ket\psi\,.
\label{eqn:schrodinger-rotated}
\eeq
In terms of the BH-aligned frame wavefunction $\ket{c}$, this becomes
\beq
\frac{\partial}{\partial t}\ket c=-i\bigl(\mathcal H+\bar V-i\mathcal R^\dagger\dot{\mathcal R}\bigr)\ket c\,.
\label{eqn:schrodinger-transformation}
\eeq
The last term in \eqref{eqn:schrodinger-transformation} is the fictitious force due to the rotation of the $xyz$ frame. We can express it in terms of $\vec\Omega$ considering an infinitesimal time interval $\dd t$,
\beq
\mathcal R(t+\dd t)=\mathcal R(t)\bigl[1-i\vec\Omega\cdot\vec L\dd t\bigr]
\implies
i\mathcal R^\dagger\dot{\mathcal R}=\vec\Omega\cdot\vec L\,.
\label{eqn:rotation-identity}
\eeq
Substituting this identity in \eqref{eqn:schrodinger-transformation} and expanding the Schrödinger equation in the basis of BH-frame eigenstates, we obtain
\beq
\dot c_a=-i\sum_b\Bigl[(E_a+i\Gamma_a)\delta_{ab}+\bar V_{ab}
-(\vec\Omega\cdot\vec L)_{ab}\Bigr]c_b\,.
\label{eqn:schrodinger-components}
\eeq

\vskip 4pt
The same derivation applies to the orbit-aligned frame, in which case we have
\beq
\frac{\partial}{\partial t}\ket{c'}
=-i\Bigl[\mathcal H'+\bar V'-(\vec\Omega'\cdot\vec L)\Bigr]\ket{c'}\,,
\label{eqn:schrodinger-cprime}
\eeq
where $\mathcal H'=d^\dagger\mathcal H d$ and $\bar V'=d^\dagger\bar Vd$. Note that $\mathcal H'$ is generally not diagonal in the basis $\ket{a'}$ introduced in Sec.~\ref{sec:setup}. Equations~\eqref{eqn:schrodinger-components} and~\eqref{eqn:schrodinger-cprime} describe the evolution of the cloud in different coordinates. To close either equation, we now need to compute the backreaction of the cloud's evolution on the BH and the orbit, which determine $\vec\Omega$ and $\vec\Omega'$.

\subsection{Backreaction on the black hole and on the orbit}
\label{sec:BHback}

Differentiating the formula for the cloud's mass in \eqref{eqn:cloud-observables} and using \eqref{eqn:schrodinger-transformation}, we get
\beq
\dot M\ped{c}=2M\ped{c,0}\braket{c|\Im\mathcal H|c}
=2M\ped{c,0}\sum_a\Gamma_a|c_a|^2\,.
\label{eqn:dotMc}
\eeq
The mass gained (lost) by the cloud is extracted from (absorbed by) the BH. Conservation of mass then requires
\beq
\dot M=-\dot M\ped{c}=-2M\ped{c,0}\sum_a\Gamma_a|c_a|^2\,.
\label{eqn:dotM}
\eeq

\vskip 4pt
Let us now compute the evolution of the cloud's spin in the freely falling frame, where we have $\vec S\ped{c}=(M\ped{c,0}/\mu)\braket{\psi|\vec L\ped{ff}|\psi}$, with $\vec L\ped{ff}$ the angular momentum operator in the freely falling frame. After taking the derivative, we can use the Schrödinger equation \eqref{eqn:schrodinger-rotated} and express the result in terms of BH-frame states as\footnote{Square (curly) brackets denote commutators (anticommutators).}
\beq
\dot{\vec S}\ped{c}\big|_{\mathrm{ff}}
=\frac{M\ped{c,0}}\mu\Big[
\underbrace{i\braket{c|[\Re\mathcal H,\vec L]|c}}_{\text{Lense--Thirring}}
+\underbrace{\braket{c|\{\Im\mathcal H,\vec L\}|c}}_{\text{horizon flux}}+\underbrace{i\braket{c|[\bar V,\vec L]|c}}_{\text{tidal}}\Big]\,.
\label{eqn:cloud-spin-fluxes}
\eeq
We introduced the subscript ``ff'' to remind the reader that, even though the right-hand side is expressed in BH-frame quantities for convenience, the derivative $\dot{\vec S}\ped{c}$ is computed in the freely falling frame.

\vskip 4pt
The terms appearing in \eqref{eqn:cloud-spin-fluxes} admit simple physical interpretations. The first is the torque due to the Lense--Thirring interaction \eqref{eqn:hyperfine}, as the spin-independent part of $\Re\mathcal H$ commutes with $\vec L$. The second is the angular momentum flux through the horizon accompanying the mass flux \eqref{eqn:dotMc}. Collectively, these give the torque $\vec\tau$ exerted by the cloud on the BH, which is responsible for the backreaction on the BH's spin,
\beq
\vec\tau=\dot{\vec S}|\ped{ff}=-\frac{M\ped{c,0}}\mu\sum_{a,b}
\bigl[i(E_a-E_b)+(\Gamma_a+\Gamma_b)\bigr]\vec L_{ab}c_a^*c_b\,.
\label{eqn:torques}
\eeq
The third and final term is the companion's tidal torque $\vec\tau_*$ on the cloud,
\beq
\vec\tau_*=\frac{iM\ped{c,0}}\mu\sum_{a,b,i}
\bigl(\bar V_{ai}\vec L_{ib}-\vec L_{ai}\bar V_{ib}\bigr)c_a^*c_b\,.
\label{eqn:tidal-torques}
\eeq
By Newton's third law, its opposite gives the tidal backreaction on the orbit.

\vskip 4pt
We have so far computed these derivatives in the freely falling frame of the gravitational atom. In a fixed external frame, the gravitational atom is also subject to geodetic precession \eqref{eqn:geodetic}: this can be described by a Hamiltonian spin-orbit coupling $H\ped{dS}=\gamma\ped{dS}\,\vec L_*\cdot(\vec S+\vec S\ped{c})$, which gives $\dot{\vec S}|\ped{dS}=i[H\ped{dS},\vec S]=\gamma\ped{dS}\vec L_*\times\vec S$, and similarly for $\vec S\ped{c}$ and $\vec L_*$. In the external frame, we can thus add up the two contributions, finally writing 
\begin{align}
\dot{\vec S}&=\vec\tau+\gamma\ped{dS}\vec L_*\times\vec S\,,
\label{eqn:dotS}\\
\dot{\vec S}\ped{c}&=-\vec\tau+\vec\tau_*+\gamma\ped{dS}\vec L_*\times\vec S\ped{c}\,,
\label{eqn:dotSc}\\
\dot{\vec L}_*&=-\vec\tau_*+\gamma\ped{dS}(\vec S+\vec S\ped{c})\times\vec L_*\,,
\label{eqn:dotL}
\end{align}
where we have imposed that the total angular momentum
\beq
\vec J\equiv\vec S+\vec S\ped{c}+\vec L_*
\eeq
remains constant, and we have neglected the companion's spin and higher-order relativistic spin couplings.

\subsection{Conserved quantities}
\label{sec:integrals}

In Sec.~\ref{sec:BHback} we used the conservation of the total mass and angular momentum (in a fixed external frame),
\begin{align}
M+M\ped{c}&=\text{const.}\,,\\
\vec S+\vec S\ped{c}+\vec L_*&=\text{const.}\,.
\end{align}
The former can also be conveniently written as a relation between the fine structure constant and the cloud-to-BH mass ratio,
\beq
\alpha=\frac{1+X_0}{1+X}\alpha_0\,,
\label{eqn:dotalpha-reduced}
\eeq
where $\alpha_0$ and $X_0$ are the values of $\alpha$ and $X=M\ped{c}/M$ at a reference initial time. The mass ratio $q=M_*/(M+M\ped{c})$ also remains constant.

\vskip 4pt
The magnitude of the orbital angular momentum is also conserved. Indeed, the orbit-averaged quadrupole potential \eqref{eqn:Vbar-quad} is axisymmetric about $\hat L_*$, so $[\bar V,\vec L\cdot\hat L_*]=0$. By Eq.~\eqref{eqn:tidal-torques}, this means that the tidal torque has no component along the orbital axis, $\vec\tau_*\cdot\hat L_*=0$. The geodetic term in \eqref{eqn:dotL} is likewise perpendicular to $\vec L_*$. Projecting \eqref{eqn:dotL} along $\hat L_*$ then gives
\beq
\dot L_*=\dot{\vec L}_*\cdot\hat L_*=0\,.
\label{eqn:Lcons}
\eeq
The backreaction on the orbit can thus change the direction of $\vec L_*$, but not its magnitude. Additionally, the semi-major axis and eccentricity are conserved\footnote{Of course, these statements cease to apply when additional effects are retained, such as gravitational radiation.} individually, because the orbit-averaged Hamiltonian is (by definition) independent of the mean anomaly, whose conjugate momentum is $L_*/\sqrt{1-e^2}$.

\vskip 4pt
Since both $J$ and $L_*$ are constant, expanding $J^2=|\vec S+\vec S\ped{c}+\vec L_*|^2$ gives a useful scalar conservation law,
\beq
K\equiv\frac{J^2-L_*^2}{2L_*}
=(\vec S+\vec S\ped{c})\cdot\hat L_*
+\frac{|\vec S+\vec S\ped{c}|^2}{2L_*}=\mathrm{const}\,.
\label{eqn:Scons}
\eeq
Equation~\eqref{eqn:Scons} will be useful in Secs.~\ref{sec:locked} and~\ref{sec:unlocked} to relate the evolution of the obliquity to that of the BH and cloud spins.

\subsection{Angular velocities of the frames}
\label{sec:gauge}

Now that we have derived the torques, we can calculate $\vec\Omega$ and $\vec\Omega'$ and feed them back in \eqref{eqn:schrodinger-components} and \eqref{eqn:schrodinger-cprime} to obtain a closed system. For the BH spin direction $\hat S=\hat z$, from \eqref{eqn:torques} we have
\beq
\dot{\hat S}|\ped{ff}
=\frac{\tau_x\hat x+\tau_y\hat y}{S}
=\vec\Omega\times\hat z\,.
\label{eqn:spin-axis-kinematics}
\eeq
This fixes
\beq
\Omega_x=-\frac{\tau_y}{S}\,,\qquad
\Omega_y=\frac{\tau_x}{S}\,,
\label{eqn:Omega}
\eeq
but it leaves $\Omega_z$ undetermined. To find it, notice that $\dot{\vec L}_*=\dot{\vec L}_*|\ped{ff}+\gamma\ped{dS}\vec L_*\times\vec L_*=\dot{\vec L}_*|\ped{ff}$. Dividing \eqref{eqn:dotL} by the constant $L_*$, we thus have
\beq
\dot{\hat L}_*|\ped{ff}
=-\frac{\vec\tau_*}{L_*}
+\gamma\ped{dS}(\vec S+\vec S\ped{c})\times\hat L_*\,.
\label{eqn:orbit-axis-dynamics}
\eeq
But from the definition \eqref{eqn:axes},
\beq
\begin{split}
\dot{\hat L}_*|\ped{ff}&=\dot\beta\cos\beta\,\hat x+\sin\beta\,\dot{\hat x}-\dot\beta\sin\beta\,\hat z+\cos\beta\,\dot{\hat z}\\
&=\vec\Omega\times\hat L_*+\dot\beta\,\hat x'\,.
\label{eqn:orbit-axis-kinematics}
\end{split}
\eeq
Equating the $y$ components of \eqref{eqn:orbit-axis-dynamics} and \eqref{eqn:orbit-axis-kinematics} determines $\Omega_z$,
\beq
\Omega_z=-\frac{\cot\beta\,\tau_y}{S}
-\frac{\tau_{*,y}}{L_*\sin\beta}
+\gamma\ped{dS}\bigl(S+S_{\text{c},z}-\cot\beta\,S_{\text{c},x}\bigr)\,.
\label{eqn:Omega-z}
\eeq
We see that the first term in \eqref{eqn:Omega-z} is present even if $\hat L_*$ is kept fixed. The second comes from the orbital backreaction induced by the tidal torque, and the third from the orbital backreaction to the geodetic precession of the gravitational atom (i.e., the effect of spin-orbit coupling on the direction of $\hat L_*$).

\vskip 4pt
The projection along $\hat x'$ determines instead the evolution of the obliquity. From $\vec\tau_*\cdot\hat L_*=0$ we have $\vec\tau_*\cdot\hat x'=-\tau_{*,z}/\sin\beta$, hence
\beq
\dot\beta=-\frac{\tau_x}S+\frac{\tau_{*,z}}{L_*\sin\beta}
+\gamma\ped{dS}S_{\text{c},y}\,.
\label{eqn:obliquity-kinematics}
\eeq

\vskip 4pt
The orbit-aligned frame is rotated by $\beta$ relative to the BH-aligned one about their common $y$ axis, so its angular velocity is $\vec\Omega'=\vec\Omega+\dot\beta\,\hat y$. Projecting this vector onto the primed axes gives
\begin{align}
\label{eqn:Omega-x-prime}
\Omega'_{x'}&=-\frac{\tau_y}{S}\cos\beta-\Omega_z\sin\beta\,,\\
\label{eqn:Omega-y-prime}
\Omega'_{y'}&=\frac{\tau_x}{S}+\dot\beta\,,\\
\label{eqn:Omega-z-prime}
\Omega'_{z'}&=-\frac{\tau_y}{S}\sin\beta+\Omega_z\cos\beta\,.
\end{align}
Both $\vec\Omega$ and $\vec\Omega'$ are now expressed in terms of the cloud's wavefunction (through the torques) and the angular momenta of BH, cloud, and orbit.

\subsection{Azimuth}
\label{sec:azimuth}

It is also useful to keep track of the absolute orientation of the gravitational atom relative to fixed external axes. We define the azimuth $\varphi$ of $\hat S$ about the constant $\hat J$, measured from a fixed direction perpendicular to $\hat J$. To find its rate of change, we can write
\beq
\dd\hat S=\hat J\times\hat S\dd\varphi+\big[\text{change of }\hat S\text{ orthogonal to }\hat J\times\hat S\big]\,.
\eeq
Projecting on $\hat J\times\hat S$, we get
\beq
\dot\varphi=\frac{\hat J\cdot(\hat S\times\dot{\hat S})}
{1-(\hat J\cdot\hat S)^2}\,.
\label{eqn:azimuth-rate}
\eeq
This derivative is taken in the fixed external frame. Using \eqref{eqn:dotS} and simplifying, we can write it as
\beq
\dot\varphi=\frac{J}S\,
\frac{S_{\text{c},y}\tau_x+
(L_*\sin\beta+S_{\text{c},x})
\big(\gamma\ped{dS}L_*S\sin\beta-\tau_y\big)}
{(L_*\sin\beta+S_{\text{c},x})^2+S_{\text{c},y}^2}\,.
\label{eqn:azimuth-rate-explicit}
\eeq
The azimuth itself does not enter the equation of motion of any other dynamical quantity, but, as we will see, it will be informative to check its evolution through \eqref{eqn:azimuth-rate-explicit}.

\subsection{Master equations}
\label{sec:master-equations}

We can now collect all coupled equations for the evolution of the cloud, BH, and orbit,
\begin{align}
\dot c_a&=(-iE_a+\Gamma_a+im_a\Omega_z)c_a-i\sum_b\bigg[
\bar V_{ab}+\frac{(\tau_yL_x-\tau_xL_y)_{ab}}{S}\bigg]c_b\,,
\label{eqn:master-ca}\\
\dot c_{a'}&=-i\sum_{b'}\bigg[
\mathcal H_{a'b'}+\bar V_{a'b'}
-\vec\Omega'\cdot\vec L_{a'b'}\bigg]c_{b'}\,,
\label{eqn:master-cprime}\\
\dot M&=-2M\ped{c,0}\sum_a\Gamma_a|c_a|^2\,,
\label{eqn:master-Mdot}\\
\dot S&=-\frac{2M\ped{c,0}}\mu\sum_a m_a\Gamma_a|c_a|^2\,,
\label{eqn:master-Sdot}\\
\dot\beta&=-\frac{\tau_x}{S}+\frac{\tau_{*,z}}{L_*\sin\beta}
+\gamma\ped{dS}S_{\text{c},y}\,,
\label{eqn:master-beta}\\
\dot\varphi&=\frac{J}{S}\,
\frac{S_{\text{c},y}\tau_x+(L_*\sin\beta+S_{\text{c},x})
\bigl(\gamma\ped{dS}L_*S\sin\beta-\tau_y\bigr)}
{(L_*\sin\beta+S_{\text{c},x})^2+S_{\text{c},y}^2}\,.
\label{eqn:master-varphi}
\end{align}
These form a closed nonlinear system that is amenable to analytical and numerical integration. Some of the quantities appearing on the right-hand side depend themselves on the wavefunction coefficients. In particular,  $\tau_x$ and $\tau_y$ are given in \eqref{eqn:torques}, $\tau_{*,z}$ in \eqref{eqn:tidal-torques}, $\Omega_z$ in \eqref{eqn:Omega-z}, $\vec\Omega'$ in \eqref{eqn:Omega-x-prime}--\eqref{eqn:Omega-z-prime}, and $S_{\text{c},x}$ and $S_{\text{c},y}$ in \eqref{eqn:cloud-observables}.

\vskip 4pt
Of course, Eq.~\eqref{eqn:master-cprime} is redundant, as it expresses the same Schrödinger equation as \eqref{eqn:master-ca} but in a different frame. It will nevertheless be useful to have at our disposal both forms of the equation and pick the most convenient one depending on the regime. In the next two sections, we will be concerned with analytically reducing the master equations \eqref{eqn:master-ca}--\eqref{eqn:master-varphi} to a simpler system.

\section{Locked co-precession}
\label{sec:locked}

In this section we consider the regime $T\sped{lt}\ll T\ped{tidal}$, which we call \emph{locked co-precession}. The Lense--Thirring coupling then dominates over the companion's tide and locks the cloud to the BH spin, so that the BH-cloud pair precesses as a whole about the total angular momentum. In Sec.~\ref{sec:adiabatic-elimination}, we use adiabatic elimination to compute the populations of the decaying states which the tide mixes with the dominant state of the cloud, and then evaluate self-consistently the torques generated by this mixing in Sec.~\ref{sec:torques}. Under the assumption of threshold drift, discussed in Sec.~\ref{sec:drift}, the evolution reduces to a one-dimensional flow for the cloud-to-BH mass ratio, which we obtain in Sec.~\ref{sec:reduced-system}. In Sec.~\ref{sec:final_locked}, we determine the final state, either spin-orbit (anti-)alignment or cloud exhaustion, directly from conservation laws. Finally, in Sec.~\ref{sec:locked-timescales}, we derive the alignment timescale and check the consistency of our approximations.

\subsection{Adiabatic elimination}
\label{sec:adiabatic-elimination}

To make the master equations \eqref{eqn:master-ca}--\eqref{eqn:master-varphi} tractable when $T\sped{lt}\ll T\ped{tidal}$, we make a physical assumption: that the cloud remains close to a single eigenstate in the BH frame, such as $\ket{a}=\ket{211}$ or $\ket{322}$. As already argued in Sec.~\ref{sec:timescales}, this can be interpreted as the consequence of the companion's tidal torque being only a weak perturbation compared to the internal BH-cloud interaction. We will verify a posteriori the consistency of this approximation. Consequently, the cloud's mass and spin will be approximately $M\ped{c}\approx M\ped{c,0}\abs{c_a}^2$ and $\vec S\ped{c}\approx m_aM\ped{c}\hat S/\mu$.

\vskip 4pt
It is useful to factor out the phase evolution of the dominant state $\ket{a}$ from the amplitudes of any state $\ket{b}$. We define
\beq
\tilde c_b\equiv c_b\exp\biggl[i\int_0^t\big(E_a-m_a\Omega_z\big)\dd t'\biggr]\,.
\label{eqn:tildecb}
\eeq
Retaining only the dominant state $\ket{a}$ in the sum in \eqref{eqn:master-ca}, the equation of motion of any state $\ket{b}\ne\ket{a}$ reads
\beq
\dot{\tilde c}_b=\big[i\big(E_a-E_b-(m_a-m_b)\Omega_z\big)+\Gamma_b\big]\tilde c_b-i\mathcal V_{ba}\tilde c_a\,,\qquad\mathcal V_{ba}\equiv\bar V_{ba}+\frac{(\tau_yL_x-\tau_xL_y)_{ba}}S\,.
\label{eqn:mathcal-V}
\eeq
As long as $\tilde c_a$ evolves on a timescale longer than the inverse energy difference appearing in \eqref{eqn:mathcal-V}, $(E_a-E_b-(m_a-m_b)\Omega_z)^{-1}$, any subdominant state with $\Gamma_b<0$ will relax to a quasi-steady-state solution found by setting $\dot{\tilde c}_b=0$. This gives
\beq
c_b\approx\frac{\mathcal V_{ba}}{E_a-E_b-(m_a-m_b)\Omega_z-i\Gamma_b}\,c_a\,.
\label{eqn:adiabatic-elimination}
\eeq
This approximation is called adiabatic elimination \cite{Takahashi:2023flk,Tomaselli:2025jfo}. However, different from other cases found in the literature, Eq.~\eqref{eqn:adiabatic-elimination} is still an implicit relation for $c_b$, because $\mathcal V_{ba}$ and $\Omega_z$ depend on the torques $\vec\tau$ and $\vec\tau_*$, which are themselves a function of the amplitudes $c_a$ and $c_b$. We complete the self-consistent solution next.

\subsection{Evaluating the torques}
\label{sec:torques}

The torque $\vec\tau$, given in \eqref{eqn:torques}, only receives nonzero contributions from adjacent states, with $\abs{\Delta m}\le1$. The leading terms in the sum are thus the ones involving at least one wavefunction coefficient of the dominant state $\ket{a}$, namely $c_a^*c_{b_1}$ and $c_{b_1}^*c_a$, where we remind the reader that $\ket{b_i}=\ket{n_a\ell_a,\ell_a-i}$. From \eqref{eqn:Lxy-matrix-elements}, the matrix elements of the angular momentum between these two states are
\beq
(L_x)_{ab_1}=(L_x)_{b_1a}=\sqrt{\frac{m_a}2}\,,\qquad
(L_y)_{ab_1}=-(L_y)_{b_1a}=-i\sqrt{\frac{m_a}2}\,,
\label{eqn:Lx-Ly-ab1}
\eeq
hence
\begin{align}
\tau_x&=-\frac{M\ped{c,0}}\mu\sqrt{2m_a}\Big[-(E_a-E_{b_1})\Im(c_a^*c_{b_1})+(\Gamma_a+\Gamma_{b_1})\Re(c_a^*c_{b_1})\Big]\,,\label{eqn:tau_x}\\
\tau_y&=-\frac{M\ped{c,0}}\mu\sqrt{2m_a}\Big[(E_a-E_{b_1})\Re(c_a^*c_{b_1})+(\Gamma_a+\Gamma_{b_1})\Im(c_a^*c_{b_1})\Big]\,.
\label{eqn:tau_y}
\end{align}
Neglecting $\Gamma_a$ and $\Gamma_{b_1}$ with respect to $E_a-E_{b_1}$ (that is, $T\sped{lt}\ll T\ped{abs}$), we can combine \eqref{eqn:tau_x} and \eqref{eqn:tau_y} into
\beq
\tau_y-i\tau_x\approx-\frac{M\ped{c,0}}\mu\sqrt{2m_a}\,(E_a-E_{b_1})c_a^*c_{b_1}\,.
\label{eqn:locked-torque-combination}
\eeq
Given the matrix elements \eqref{eqn:Lx-Ly-ab1}, this is precisely the combination entering $\mathcal V_{b_1a}$, defined in \eqref{eqn:mathcal-V}. We can thus write
\beq
\mathcal V_{b_1a}c_a
=\bar V_{b_1a}c_a+\sqrt{\frac{m_a}2}\,\frac{\tau_y-i\tau_x}{S}c_a
\approx\bar V_{b_1a}c_a-\frac{S\ped c}{S}(E_a-E_{b_1})c_{b_1}\,.
\label{eqn:locked-feedback}
\eeq
On the other hand, we have $\mathcal V_{b_2a}=\bar V_{b_2a}$ because the torque terms only connect states with $\abs{\Delta m}\le1$. Substituting \eqref{eqn:locked-feedback} into \eqref{eqn:adiabatic-elimination}, using again $T\sped{lt}\ll T\ped{abs}$, and solving for $c_{b}$ gives
\begin{align}
c_{b_1}\approx-\frac{T\sped{lt}}{T\ped{tidal}}\frac{\mathcal C_1\sin\beta\cos\beta}{1+S\ped c/S-\Omega_zT\sped{lt}}\,c_a\,,\qquad c_{b_2}\approx-\frac{T\sped{lt}}{2T\ped{tidal}}\frac{\mathcal C_2\sin^2\beta}{1-\Omega_zT\sped{lt}}\,c_a\,.
\label{eqn:locked-populations}
\end{align}
This shows that $c_a^*c_{b_1}$ is real, so \eqref{eqn:tau_x} with the $\Gamma$ terms neglected gives $\tau_x=0$ and
\beq
\tau_y\approx\mathcal C_1\sqrt{\frac2{m_a}}\,\frac{S\ped{c}}{T\ped{tidal}}\frac{\sin\beta\cos\beta}{1+S\ped c/S-\Omega_zT\sped{lt}}\,,
\label{eqn:tau_y_simple}
\eeq
at leading order in $T\sped{lt}/T\ped{abs}$ and $T\sped{lt}/T\ped{tidal}$. The coefficients $c_{b_i}$ for $i\ge3$ are suppressed by higher powers of $\bar V$, because a quadrupolar perturbation only couples states with $\abs{\Delta m}\le2$.

\vskip 4pt
Two comments are in order. First, because of the dependence of $\Omega_z$ on $\tau_y$, as given in \eqref{eqn:Omega-z}, it may look like \eqref{eqn:locked-populations} and \eqref{eqn:tau_y_simple} are still implicit expressions. However, $\tau_y$ is naturally of order $\mathcal O(S\ped{c}/T\ped{tidal})$, or smaller if the backreaction is strong enough for $\Omega_z$ to dominate the denominator. This introduces a relative correction to \eqref{eqn:locked-populations} of order at most $\mathcal O(T\sped{lt}/T\ped{tidal})$. On the other hand, the second and third terms in $\Omega_z$ arise from the orbital backreaction and geodetic precession, whose magnitudes do not depend on $T\sped{lt}/T\ped{tidal}$ alone, but also on the value of $L_*$, so they are not necessarily negligible. We can thus discard the dependence of $\Omega_z$ on $\tau_y$ in \eqref{eqn:Omega-z}. To leading order, from~\eqref{eqn:tidal-torques} we have
\beq
\tau_{*,y}\approx\sqrt{\frac2{m_a}}\,\frac{\mathcal C_1S\ped c}{T\ped{tidal}}\sin\beta\cos\beta\,,
\label{eqn:locked-tidal-torque}
\eeq
and thus
\beq
\Omega_z\approx-\sqrt{\frac2{m_a}}\,\frac{\mathcal C_1}{T\ped{tidal}}\frac{S\ped c}{L_*}\cos\beta+\gamma\ped{dS}(S+S\ped{c})\,.
\label{eqn:Omega-z-locked}
\eeq

\vskip 4pt
Second, the factors of $1/(1+S\ped{c}/S)$ in \eqref{eqn:locked-populations} and \eqref{eqn:tau_y_simple} enjoy a simple physical interpretation. Suppose that the orbital plane remains approximately fixed, so that $\Omega_z$ is negligible. Because the BH and the cloud are locked together, the torque exerted by the companion makes the whole pair precess at once, and the angular momentum of the gravitational atom exceeds that of the BH by a factor of $1+S\ped{c}/S$. The torque is defined in \eqref{eqn:torques} as acting on the spin of the BH only, so the appearance of this corrective factor in \eqref{eqn:locked-populations} and \eqref{eqn:tau_y_simple} precisely adjusts for that.

\subsection{Threshold drift}
\label{sec:drift}

To complete the analytical reduction of the master equations \eqref{eqn:master-ca}--\eqref{eqn:master-varphi}, we need one last key insight. A BH that grows a superradiant cloud in isolation saturates the instability on the \emph{superradiant threshold}, defined by $\Gamma_a=0$. This is a stable configuration: if the system is perturbed such that the BH spin falls just below the threshold, the grown state $\ket{a}$ will now be decaying, donating some angular momentum back to the BH, until the threshold is reached again. Conversely, if the system goes above the threshold, superradiant instability will kick in once again and reduce the BH spin. While the cloud loses (gains) mass in the former (latter) case, the net effect on the BH mass is \emph{always} an increase, which can be seen as a consequence of Hawking's area law determining which direction along the threshold is thermodynamically allowed. When a gravitational atom is perturbed, it therefore undergoes a \emph{threshold drift} towards larger values of $\alpha=\mu M$. This phenomenon has been observed in~\cite{Brito:2014wla} in the context of BH accretion, while~\cite{Hui:2022sri} studied the reduced dynamics and applied it to the growth of slower superradiant states.

\vskip 4pt
Threshold drift relies on the fundamental assumption that, \emph{away} from the threshold, the growth or decay of state $\ket{a}$ operates on a much faster timescale than the change in BH mass and spin induced by the external perturbation (in other words, $\Gamma_a$ varies ``steeply'' as one moves away from $\Gamma_a=0$). As a consequence, a tiny deviation from the threshold is sufficient to make the cloud's mass decay rate $-2M\ped{c,0}\Gamma_a\abs{c_a}^2$ large enough to balance the external perturbation on $M$ and keep the BH near the threshold. A massive (less massive) cloud needs a smaller (larger) deviation from the threshold to keep up. Whenever the cloud becomes depleted beyond a critical amount, its decay can no longer change the BH parameters fast enough to compensate for the external perturbation, and the threshold drift breaks. The subsequent trajectory of the BH in the mass-spin plane is determined by the external perturbation only. In this section, we will assume that the hypothesis of threshold drift holds---we will confirm in Sec.~\ref{sec:locked-timescales} that the resulting dynamics is self-consistent.

\vskip 4pt
For the three states $\ket{a}$, $\ket{b_1}$, and $\ket{b_2}$ involved in our discussions in Secs.~\ref{sec:adiabatic-elimination} and \ref{sec:torques}, the master equations \eqref{eqn:master-Mdot}--\eqref{eqn:master-Sdot} read\footnote{The sums appearing in \eqref{eqn:dotM-adiabatic}--\eqref{eqn:dotS-adiabatic} should in principle be extended to all other states. As explained below \eqref{eqn:Vbar}, however, we restrict our analysis to a given hyperfine multiplet. This is a good approximation for $\ket{a}=\ket{211}$ because other states with nonzero matrix elements have $n\ge3$, e.g., $\ket{31\,\minus1}$ and $\ket{433}$. Due to their large energy splitting with $\ket{211}$, their population is greatly suppressed according to \eqref{eqn:adiabatic-elimination}. The approximation is however not necessarily good for $\ket{a}=\ket{322}$, as $\ket{300}$ enters the dynamics with a non-negligible matrix element and an extremely large decay rate.}
\begin{align}
\dot M&=-2M\ped c\bigg(\Gamma_a+\Gamma_{b_1}\Big|\frac{c_{b_1}}{c_a}\Big|^2+\Gamma_{b_2}\Big|\frac{c_{b_2}}{c_a}\Big|^2\bigg)\,,\label{eqn:dotM-adiabatic}\\
\dot S&=-\frac{2M\ped c}\mu\bigg(m_a\Gamma_a+(m_a-1)\Gamma_{b_1}\Big|\frac{c_{b_1}}{c_a}\Big|^2+(m_a-2)\Gamma_{b_2}\Big|\frac{c_{b_2}}{c_a}\Big|^2\bigg)\,.
\label{eqn:dotS-adiabatic}
\end{align}
Under the hypothesis of threshold drift, instead of computing $\Gamma_a$ directly (e.g., using \eqref{eqn:Gamma_nlm}), we can consider $\Gamma_a$ as a new variable and eliminate it by enforcing the threshold condition,
\beq
\dot S=\frac{\dd S}{\dd M}\bigg|_{\Gamma_a=0}\dot M\,.
\eeq
We can then combine \eqref{eqn:dotM-adiabatic} and \eqref{eqn:dotS-adiabatic} to get
\beq
\dot M=-\frac{2M\ped c}{m_a}\bigg(1-\frac\mu{m_a}\frac{\dd S}{\dd M}\bigg|_{\Gamma_a=0}\bigg)^{-1}\bigg[\Gamma_{b_1}\Big|\frac{c_{b_1}}{c_a}\Big|^2+2\Gamma_{b_2}\Big|\frac{c_{b_2}}{c_a}\Big|^2\bigg]_{\Gamma_a=0}\,,
\label{eqn:dotM-threshold}
\eeq
where the prefactor can be approximated with \eqref{eqn:threshold} as
\beq
\bigg(1-\frac\mu{m_a}\frac{\dd S}{\dd M}\bigg|_{\Gamma_a=0}\bigg)^{-1}\approx\frac{(m_a^2+4\alpha^2)^2}{m_a^2(m_a^2-4\alpha^2)}\,.
\label{eqn:kappa}
\eeq
Equation \eqref{eqn:dotM-threshold} gives the total BH mass change during the threshold drift, which takes into account the absorption of the dominant state $\ket{a}$ and is triggered by the decay of states $\ket{b_1}$ and $\ket{b_2}$.

\subsection{Reduced system of equations}
\label{sec:reduced-system}

When combined with the adiabatic elimination \eqref{eqn:locked-populations} and the mass conservation \eqref{eqn:dotalpha-reduced} formulae, Eq.~\eqref{eqn:dotM-threshold} gives a one-dimensional closed-form expression for the evolution of the cloud-to-BH mass ratio $X=M\ped{c}/M$,
\beq
\dot X=2X(1+X)\frac{(m_a^2+4\alpha^2)^2}{m_a^3(m_a^2-4\alpha^2)}\bigg(\frac{T\sped{lt}}{T\ped{tidal}}\bigg)^2\Bigg[\frac{\mathcal C_1^2\Gamma_{b_1}\sin^2\beta\cos^2\beta}{(1+S\ped c/S-\Omega_zT\sped{lt})^2}+\frac{\mathcal C_2^2\Gamma_{b_2}\sin^4\beta}{2(1-\Omega_zT\sped{lt})^2}\Bigg]_{\Gamma_a=0}\,.
\label{eqn:dotX-reduced}
\eeq
The widths $\Gamma_{b_1}$ and $\Gamma_{b_2}$ are necessarily negative on the $\Gamma_a=0$ threshold, so the cloud's mass decreases ($\dot X<0$) whenever $\beta\ne0$ or $\pi$.

\vskip 4pt
Remarkably, all the remaining variables can be determined algebraically from $X$. As already mentioned, mass conservation fixes $\alpha(X)$ through \eqref{eqn:dotalpha-reduced}, and the threshold condition determines the BH spin $S(X)$ via \eqref{eqn:threshold}. To find the obliquity $\beta$, we can use angular momentum conservation, specifically the invariant \eqref{eqn:Scons}, which in the locked case reads $K=(S+S\ped c)\cos\beta+(S+S\ped c)^2/(2L_*)$, where we remind the reader that $S\ped{c}=m_aXM^2/\alpha$. Inverting the equation gives
\beq
\cos\beta=\frac{(S_0+S\ped{c,0})\cos\beta_0+\big[(S_0+S\ped{c,0})^2-(S+S\ped c)^2\big]/(2L_*)}{S+S\ped c}\,,
\label{eqn:cosbeta-reduced}
\eeq
where $S_0$ and $S\ped{c,0}=m_aM\ped{c,0}/\mu$ are the initial BH and cloud spin magnitudes. In the limit $L_*\gg S+S\ped c$, where orbital backreaction is negligible, Eq.~\eqref{eqn:cosbeta-reduced} reduces to the conservation of $(S+S\ped c)\cos\beta$.

\vskip 4pt
The rate of change of the obliquity can be found by taking the derivative of \eqref{eqn:cosbeta-reduced}, yielding
\beq
\begin{split}
\dot\beta&=\frac{m_a^2(m_a^2-4\alpha^2)}{(m_a^2+4\alpha^2)^2}\,\frac{\dot X}{X(1+X)}\,\bigg(\frac{\cot\beta}{1+S/S\ped{c}}+\frac{S\ped{c}}{L_*\sin\beta}\bigg)\\
&=\frac2{m_a}\bigg(\frac{T\sped{lt}}{T\ped{tidal}}\bigg)^2\bigg(\frac{\cot\beta}{1+S/S\ped{c}}+\frac{S\ped{c}}{L_*\sin\beta}\bigg)\Bigg[\frac{\mathcal C_1^2\Gamma_{b_1}\sin^2\beta\cos^2\beta}{(1+S\ped c/S-\Omega_zT\sped{lt})^2}+\frac{\mathcal C_2^2\Gamma_{b_2}\sin^4\beta}{2(1-\Omega_zT\sped{lt})^2}\Bigg]_{\Gamma_a=0}\,.
\end{split}
\label{eqn:dotbeta-reduced}
\eeq
We see that obliquity decreases ($\dot\beta<0$) for $\cos\beta+(S+S\ped c)/L_*>0$, and increases ($\dot\beta>0$) otherwise. In other words, while the cloud depletes, the gravitational atom's spin and the orbit's angular momentum vector progressively align, either towards the aligned ($\beta=0$) or anti-aligned ($\beta=\pi$) configuration. Note, however, that while $L_*$ is a constant, $S+S\ped{c}$ decreases as the cloud depletes, as
\beq
\dot S+\dot S\ped{c}=\frac{\dd S}{\dd M}\bigg|_{\Gamma_a=0}\dot M+\frac{m_a}\mu\dot M\ped{c}=-\frac{m_a^3(m_a^2-4\alpha^2)}{\mu(m_a^2+4\alpha^2)^2}\dot M<0\,.
\label{eqn:S+Sc-decreases}
\eeq
Thus the separatrix between $\dot\beta\gtrless0$ also evolves during the evolution.

\vskip 4pt
Finally, the azimuthal precession rate of the locked BH-cloud pair about $\vec J$ can be found by simplifying \eqref{eqn:master-varphi}. The component $S_{\text{c},y}$ can be neglected, while comparing \eqref{eqn:tau_y} with \eqref{eqn:cloud-observables} we can write $S_{\text{c},x}\approx-T\sped{lt}\tau_y$. Substituting this into \eqref{eqn:master-varphi} and using \eqref{eqn:tau_y_simple} and \eqref{eqn:Omega-z-locked}, we get
\beq
\begin{split}
\dot\varphi&=J\,\frac{\gamma\ped{dS}L_*\sin\beta-\tau_y/S}{L_*\sin\beta-T\sped{lt}\tau_y}\\
&=J\,\frac{\Omega_z}{S+S\ped{c}}\\
&=J\bigg[\gamma\ped{dS}-\sqrt{\frac2{m_a}}\frac{\mathcal C_1}{T\ped{tidal}}\frac{\cos\beta}{(1+S/S\ped c)L_*}\bigg]\,.
\end{split}
\label{eqn:dotphi-locked}
\eeq
The first term in brackets is the unavoidable geodetic precession. The second term is best interpreted in the limit of large $L_*$, where the orbital plane remains fixed and the prefactor $J$ cancels with the $L_*$ in the denominator. We then see that the locked BH-cloud pair precesses about $\vec J$ at a rate determined by the tidal torque acting on the cloud, with the factor of $1+S/S\ped{c}$ in the denominator accounting for the fact that the cloud ``drags'' the BH to precess with it.

\subsection{Final state: spin-orbit (anti-)alignment or cloud exhaustion}
\label{sec:final_locked}

In Sec.~\ref{sec:reduced-system} we showed that the evolution of the system can be reduced to a one-dimensional system. Determining the final state is even simpler, because it follows directly from conservation laws, bypassing the integration of \eqref{eqn:dotX-reduced}.

\vskip 4pt
Equations \eqref{eqn:dotalpha-reduced}, \eqref{eqn:threshold}, and \eqref{eqn:cosbeta-reduced} determine the trajectory of the system in the $\{X,\beta,\alpha\}$ parameter space and can thus be used to predict the endpoint of the evolution. From \eqref{eqn:dotX-reduced}, it is clear that there are two possible final states, where $\dot X=\dot\beta=0$: either full spin (anti-)alignment is achieved ($\sin\beta=0$) or the cloud is completely exhausted ($X=0$). In this section, we derive the condition that determines which of the two outcomes occurs, as well as the final parameters of the system, which we indicate with a subscript $f$ (e.g., the final obliquity $\beta_f$, cloud-to-BH mass ratio $X_f$, etc.).

\vskip 4pt
The magnitude of the orbital angular momentum, $L_*$, and that of the total angular momentum,
\beq
J=\abs{\vec L_*+\vec S_0+\vec S\ped{c,0}}=\sqrt{L_*^2+(S_0+S\ped{c,0})^2+2L_*(S_0+S\ped{c,0})\cos\beta_0}\,,
\label{eqn:J-locked-initial}
\eeq
where
\beq
S_0+S\ped{c,0}=\frac{m_aM_0^2}{\alpha_0}\bigg(X_0+\frac{4\alpha_0^2}{m_a^2+4\alpha_0^2}\bigg)\,,
\eeq
are conserved throughout the evolution. Unless the system is already initially aligned ($\beta_0=0$ or $\pi$), by the triangle inequality we have $\abs{J-L_*}<S_0+S\ped{c,0}<J+L_*$. As argued in \eqref{eqn:S+Sc-decreases}, the sum $S+S\ped{c}$ \emph{decreases} during threshold drift, therefore the two vectors $\vec L_*$ and $\vec S+\vec S\ped{c}$ can become collinear only when they hit the lower bound of the triangle inequality, i.e., their magnitudes satisfy $S+S\ped{c}=\abs{J-L_*}$. The lowest possible value attainable by $S+S\ped{c}$ corresponds to the cloud's exhaustion: we denote it by $S|_{X=0}$, and it equals the BH spin evaluated on the threshold at the BH mass $M_0+M\ped{c,0}=M_0(1+X_0)$. In terms of the initial data, we can write it explicitly as
\beq
S\big|_{X=0}=
\frac{4m_a\alpha_0M_0^2(1+X_0)^3}{m_a^2+4\alpha_0^2(1+X_0)^2}\,.
\label{eqn:critical-condition}
\eeq
We can thus distinguish the two outcomes as follows.

\paragraph{Spin (anti-)alignment.} If $S|_{X=0}<\abs{J-L_*}$, the system will reach spin (anti-)alignment before the cloud is exhausted. We can then further distinguish two cases. If $J>L_*$ we must have $J=L_*+S+S\ped{c}$, meaning spin-orbit alignment ($\beta_f=0$). Otherwise, if $J<L_*$, then $J=L_*-S-S\ped{c}$ and we have anti-alignment ($\beta_f=\pi$). These cases can be written as algebraic conditions on the initial obliquity,
\begin{align}
\label{eqn:spin-alignment}
\text{spin alignment:}&\qquad\cos\beta_0>\frac{S|_{X=0}}{S_0+S\ped{c,0}}-\frac{(S_0+S\ped{c,0})^2-S^2|_{X=0}}{2L_*(S_0+S\ped{c,0})}\,,\\[0pt]
\label{eqn:spin-anti-alignment}
\text{spin anti-alignment:}&\qquad\cos\beta_0<-\frac{S|_{X=0}}{S_0+S\ped{c,0}}-\frac{(S_0+S\ped{c,0})^2-S^2|_{X=0}}{2L_*(S_0+S\ped{c,0})}\,.
\end{align}
In the limit of small orbital backreaction (large $L_*$), these intervals become symmetrical, giving $\abs{\cos\beta_0}>S|_{X=0}/(S_0+S\ped{c,0})$. The gravitational atom then aligns towards $\vec L_*$ or $-\vec L_*$ depending on which one is closest to its initial spin direction. Instead, in the limit of large backreaction, (small $L_*$), all initial obliquities lead to spin alignment. The final cloud mass can be determined by solving $S_f+S_{\text{c},f}=\abs{J-L_*}$. This gives a quadratic equation with solution
\beq
X_f=\frac{1+X_0}{2\Big(1+X_0-\frac{\alpha_0\abs{J-L_*}}{m_aM_0^2}\Big)}\Bigg[1+\sqrt{1-\frac{16\alpha_0^2}{m_a^2}\bigg(1+X_0-\frac{\alpha_0\abs{J-L_*}}{m_aM_0^2}\bigg)^2}\,\Bigg]-1\,.
\label{eqn:Xf-locked}
\eeq

\paragraph{Cloud exhaustion.} If instead $S|_{X=0}>\abs{J-L_*}$, the cloud is not massive enough to survive, and it is exhausted before achieving spin (anti-)alignment. Setting $X=0$ in \eqref{eqn:cosbeta-reduced} gives the final obliquity,
\beq
\begin{split}
\cos\beta_f={}&\frac{m_a^2+4\alpha_0^2(1+X_0)^2}{4\alpha_0^2(1+X_0)^3}\bigg(X_0+\frac{4\alpha_0^2}{m_a^2+4\alpha_0^2}\bigg)\bigg[\cos\beta_0+\frac{m_aM_0^2}{2\alpha_0L_*}\bigg(X_0+\frac{4\alpha_0^2}{m_a^2+4\alpha_0^2}\bigg)\bigg]\\
&-\frac{2m_a\alpha_0M_0^2(1+X_0)^3}{L_*\big[m_a^2+4\alpha_0^2(1+X_0)^2\big]}\,.
\end{split}
\label{eqn:betaf-locked}
\eeq
Setting $\cos\beta_f=\pm1$ and using \eqref{eqn:critical-condition} recovers the expressions in \eqref{eqn:spin-alignment} and \eqref{eqn:spin-anti-alignment}.

\begin{figure}
\centering
\includegraphics[width=\textwidth]{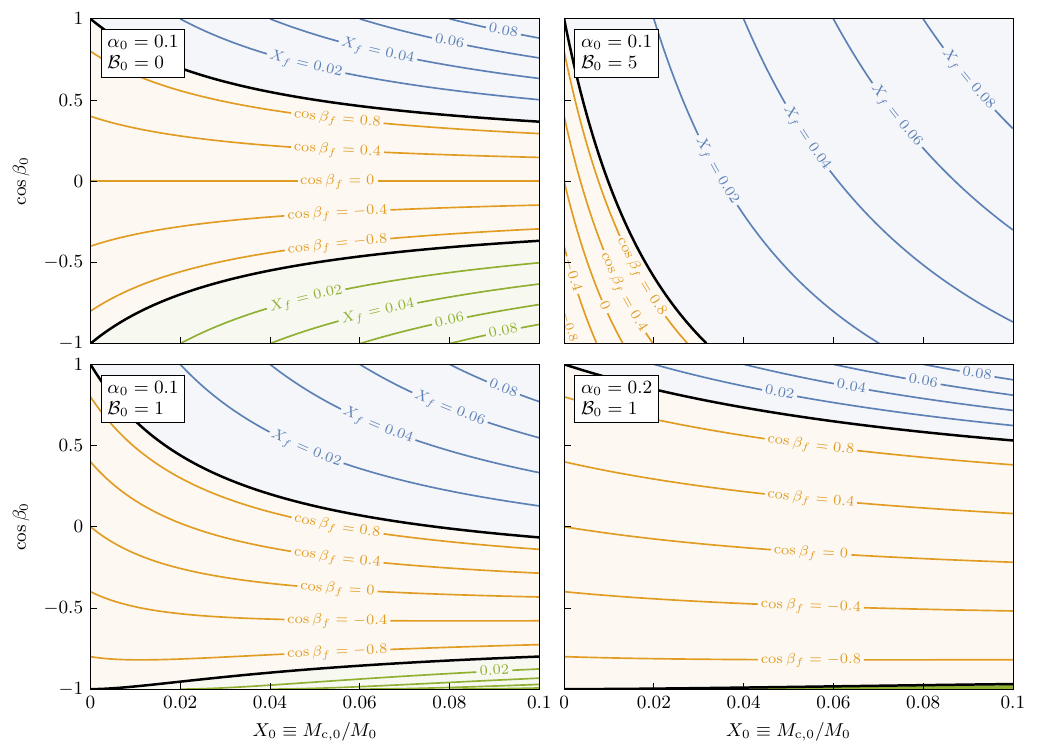}
\caption{Final state of the gravitational atom in the locked co-precession regime, as a function of the initial cloud mass fraction $X_0\equiv M\ped{c,0}/M_0$ and obliquity $\beta_0$. We vary the (initial) gravitational fine structure constant $\alpha_0$ and backreaction parameter $\mathcal B_0$, defined in \eqref{eqn:backreaction-parameter}, as indicated in the insets. All panels assume $m_a=1$. The thick black curves separate two qualitatively different outcomes. In the \textcolor{Mathematica1}{blue} and \textcolor{Mathematica3}{green} regions, the system reaches spin alignment ($\cos\beta_f=1$) or anti-alignment ($\cos\beta_f=-1$), respectively, with a residual cloud mass fraction $X_f$. In the \textcolor{Mathematica2}{orange} region, the cloud is exhausted ($X_f=0$) and the system reaches a final obliquity $\beta_f$. Note that while we cut off all panels at $X_0=0.1$, a superradiant evolution in vacuum can give at most $X_0=0.078$ for $\alpha_0=0.1$ and $X_0=0.107$ for $\alpha_0=0.2$, see \eqref{eq:X_0_max} and \cite{Herdeiro:2021znw,Hui:2022sri}.}
\label{fig:final-state}
\end{figure}

\vskip 4pt
The two outcomes are mutually exclusive. We show in Fig.~\ref{fig:final-state} the final state of the system as a function of $X_0$. We display four cases, encompassing $\alpha_0=0.1$ and $\alpha_0=0.2$, as well as different magnitudes of orbital backreaction, which we quantify through the ratio of angular momentum reservoirs in the gravitational atom and in the binary orbit,
\beq
\mathcal B_0\equiv\frac{S_0+S\ped{c,0}}{L_*}\,.
\label{eqn:backreaction-parameter}
\eeq
In the absence of backreaction ($\mathcal B_0=0$) one can see that the outcomes are symmetric with respect to $\cos\beta_0=0$, as noted earlier. Increasing $\mathcal B_0$ at fixed $\alpha_0$ and $X_0$ enlarges the region of spin alignment, reducing the portions of parameter space corresponding to spin anti-alignment and cloud exhaustion. For large enough $\mathcal B_0$, the only possible final state is spin alignment, which can be interpreted as orbital reorientation costing a negligible amount of cloud depletion. At fixed $\mathcal B_0$, a smaller $\alpha_0$ leads to a larger region where spin (anti-)alignment is achieved and the cloud survives.

\vskip 4pt
We also remind the reader that, although in Fig.~\ref{fig:final-state} we treat $X_0$ as an independent parameter and truncate its axis at $X_0=0.1$, the initial cloud mass fraction is determined by the initial BH parameters. For superradiance growth around an isolated BH, this can be at most\footnote{This formula is obtained, e.g., from Eq.~(2.18) of \cite{Hui:2022sri} assuming an initially extremal BH.}~\cite{Herdeiro:2021znw,Hui:2022sri}
\beq
X_0\le\bigg(\frac{m_a}{2\alpha_0}-1\bigg)\bigg(1-\frac1{\sqrt{1+(2\alpha_0/m_a)^2}}\bigg)\,,
\label{eq:X_0_max}
\eeq
which scales as $X_0\to \alpha_0/m_a$ for small $\alpha_0/m_a$.

\subsection{Alignment timescale}
\label{sec:locked-timescales}

The characteristic timescales over which the cloud depletes and the spin aligns with the orbit can be derived from \eqref{eqn:dotX-reduced} and \eqref{eqn:dotbeta-reduced} respectively. To obtain them, let us first expand the terms $\Omega_zT\sped{lt}$ appearing in the denominators of \eqref{eqn:dotX-reduced} and \eqref{eqn:dotbeta-reduced},
\beq
\begin{split}
\Omega_zT\sped{lt}&=-\sqrt{\frac2{m_a}}\,\mathcal C_1\frac{T\sped{lt}}{T\ped{tidal}}\frac{S\ped c}{L_*}\cos\beta+\gamma\ped{dS}T\sped{lt}(S+S\ped{c})\\
&\sim-0.0417\,X\,\bigg(\frac{0.2}\alpha\bigg)^{11}\bigg(\frac{10^3}{a/M}\bigg)^{7/2}+3.75\times10^{-4}\bigg(1+\frac{m_a^2X}{4\alpha^2}\bigg)\bigg(\frac{0.2}\alpha\bigg)^6\bigg(\frac{10^3}{a/M}\bigg)^{3}\,,
\end{split}
\label{eqn:Omegaz-Tlt-expanded}
\eeq
where we assumed $\ket{a}=\ket{211}$ and neglected factors of $\mathcal O(1)$. While the second term in \eqref{eqn:Omegaz-Tlt-expanded} is often negligible when the companion is outside the cloud ($a\gg r\ped{c}$), the first term may be either smaller or larger than unity, due to the strong dependence on $\alpha$ and $a/M$. We thus keep it in what follows.

\vskip 4pt
Close to alignment, the $\ket{b_1}$ term dominates the square brackets in \eqref{eqn:dotX-reduced} and \eqref{eqn:dotbeta-reduced}, which reduce to $\dot\beta\simeq-\sin\beta/T\ped{align}$ and (at small $\alpha$) $\dot X\simeq-X\sin^2\beta/T\ped{depl}$, where we defined the characteristic depletion and alignment timescales as
\beq
T\ped{depl}\equiv\frac{m_a}{2\mathcal C_1^2}\bigg(1+\frac{S\ped{c}}S-\Omega_zT\sped{lt}\bigg)^2T\ped{abs}\bigg(\frac{T\ped{tidal}}{T\sped{lt}}\bigg)^2\,,\qquad T\ped{align}\equiv\frac{1+S/S\ped{c}}{1+\mathcal B}\,T\ped{depl}\,.
\label{eqn:T_align}
\eeq
Here $\mathcal B\equiv(S+S\ped{c})/L_*$ is the backreaction parameter \eqref{eqn:backreaction-parameter} evaluated at the instantaneous state of the system, and $\Omega_z$ is evaluated at $\cos\beta=1$. Close to anti-alignment ($\cos\beta=-1$), the same expressions hold with $\beta\to\pi-\beta$ and $\mathcal B\to-\mathcal B$. The relevance of the $\Omega_zT\sped{lt}$ term depends on the ratio
\beq
\frac{S\ped{c}}{L_*}=\frac{m_aX\sqrt{1+q}}{\alpha q(1+X)^{3/2}}\sqrt{\frac{M}{a(1-e^2)}}\,,
\eeq
which can be considered as a proxy for the strength of orbital backreaction when $S\ped{c}/S\sim\mathcal O(1)$, which is typically the case for generic BH mass and spin at BH formation. For small backreaction, $\mathcal B\ll1$, we can neglect $\Omega_zT\sped{lt}$ and Eq.~\eqref{eqn:T_align} simplifies to
\beq
T\ped{depl}\simeq\frac{m_a}{2\mathcal C_1^2}\bigg(1+\frac{S\ped{c}}S\bigg)^2T\ped{abs}\bigg(\frac{T\ped{tidal}}{T\sped{lt}}\bigg)^2\,,\qquad T\ped{align}\simeq\bigg(1+\frac S{S\ped{c}}\bigg)T\ped{depl}\,,
\label{eqn:T_align-small}
\eeq
where $2\mathcal C_1^2/m_a=81$ for $\ket{211}$ and $729$ for $\ket{322}$.

\vskip 4pt
For large backreaction, $\mathcal B\gg1$, it is convenient to define the separation $a_\Omega$ at which the first term in \eqref{eqn:Omegaz-Tlt-expanded} equals $1+S\ped{c}/S$ in magnitude at $\cos\beta=1$, namely
\beq
\frac{a_\Omega}M\simeq\mathcal C_3\,\bigg(\frac{X}{1+S\ped{c}/S}\bigg)^{2/7}\alpha^{-22/7}(1-e^2)^{-4/7}\,,
\eeq
with $\mathcal C_3\equiv(\sqrt{2m_a}\,\mathcal C_1\alpha^7T\sped{lt}/M)^{2/7}$, that is, $\mathcal C_3\simeq2.6$ for $\ket{211}$ and $\mathcal C_3\simeq12$ for $\ket{322}$. Neglecting the geodetic precession, we then have
\beq
T\ped{depl}\simeq\bigg(\frac{S\ped{c}}{L_*}\bigg)^2\bigg[1+\bigg(\frac a{a_\Omega}\bigg)^{7/2}\bigg]^2T\ped{abs}\,,\qquad T\ped{align}\simeq\frac{L_*}{S\ped{c}}\,T\ped{depl}\,.
\label{eqn:T_align-strong}
\eeq
For $a\gg a_\Omega$, $T\ped{depl}$ is the same as in \eqref{eqn:T_align-small}, while $T\ped{align}$ is reduced by a factor of $\mathcal B$. For $a\ll a_\Omega$, instead, $T\ped{align}$ saturates to $(S\ped{c}/L_*)T\ped{abs}$, which does not depend on the strength of the tide.

\vskip 4pt
In physical units, the small-backreaction alignment timescale reads
\beq
\begin{split}
T\ped{align}\simeq{}&\frac{(1-e^2)^3(1+X)^{-2}}{q^2}\bigg(1+\frac{S\ped{c}}S\bigg)^2\bigg(1+\frac S{S\ped{c}}\bigg)\bigg(\frac{M}{10M_\odot}\bigg)\bigg(\frac{a/M}{10^3}\bigg)^6\\
&\times\begin{cases}
\displaystyle\SI{2.6e-3}{yr}\,\bigg(\frac{\alpha}{0.2}\bigg)^{10}\frac1{1+4\alpha^2}&\ket{211}\,,\\[8pt]
\displaystyle\SI{0.37}{yr}\,\bigg(\frac{\alpha}{0.2}\bigg)^6\frac{1+\alpha^2}{1-\alpha^2+\alpha^4}&\ket{322}\,.
\end{cases}
\end{split}
\label{eqn:T-align-explicit}
\eeq

While the cloud depletes and the spins align, the gravitational atom precesses around $\vec J$ at a rate set by \eqref{eqn:dotphi-locked}. Let us ignore the inevitable geodetic precession. When the backreaction is negligible (so that the prefactor $J$ cancels with $L_*$ in the denominator), we see that the precession rate is set by the Newtonian torque acting on the cloud. For strong backreaction, we have instead the same enhancement factor of $\mathcal B$ as before. Setting $\cos\beta=1$ in \eqref{eqn:dotphi-locked}, we can define the characteristic precession timescale as
\beq
T\ped{prec}\equiv\frac{1+S/S\ped{c}}{1+\mathcal B}\,\frac{T\ped{tidal}}{\sqrt{2/m_a}\,\mathcal C_1}\,.
\eeq

\vskip 4pt
Finally, we remind the reader that all results in the locked co-precession regime have been derived under the assumption that $T\sped{lt}/T\ped{tidal}\ll1$, where the explicit expression of the ratio of timescales is given in \eqref{eqn:timescales-ratio}. This underlies, e.g., the adiabatic elimination \eqref{eqn:locked-populations}.\footnote{More precisely, we need $\abs{c_{b_1}/c_a}\ll1$, which for $a\ll a_\Omega$ only requires $S\ped{c}\gg L_*$, even if $T\ped{tidal}\lesssim T\sped{lt}$.} Furthermore, in Sec.~\ref{sec:drift} we assumed the threshold drift condition, deferring to this section a consistency check of such assumption. To keep the BH close to the threshold, the backreaction due to the decay of the dominant state $\ket{a}$ must outpace the evolution rate of the system, given by $T\ped{depl}$. The characteristic timescale of such backreaction is $(S/S\ped{c})T\ped{abs}$, so for threshold drift we require $T\ped{depl}\gg(S/S\ped{c})T\ped{abs}$. When $S\ped c/S\sim\mathcal O(1)$, this is satisfied under the same hypothesis underlying the rest of the results in this section. The approximation, however, fails near cloud exhaustion, when its spin becomes very small. The ratio $T\sped{lt}/T\ped{tidal}$ also controls the misalignment of the cloud and BH spins, with the equatorial component of the cloud's spin being $S\ped{c}^\perp=(2M\ped{c,0}/\mu)(L_x)_{ab_1}c_a^*c_{b_1}\sim(T\sped{lt}/T\ped{tidal})S\ped{c}$.

\section{Unlocked alignment}
\label{sec:unlocked}

In this section we consider the regime opposite to Sec.~\ref{sec:locked}, namely $T\ped{tidal}\ll T\sped{lt}$, which we call \emph{unlocked alignment}. We will show that the cloud is in an eigenstate of the tidal interaction, with angular momentum aligned along the orbital axis rather than along the BH spin. We compute the frequencies and widths of the orbit-aligned states in Sec.~\ref{sec:diagonalization}, and identify the dominant one. Then, in Sec.~\ref{sec:unlocked-bh-backreaction}, we derive the resulting evolution of the cloud, the BH, and the obliquity. In Sec.~\ref{sec:unlocked-final}, we discuss the final states, which can sometimes no longer be predicted from conservation laws alone, and argue that one of the two anti-aligned branches is unstable. We end in Sec.~\ref{sec:unlocked-timescales} with the alignment timescale.

\subsection{Diagonalization of the tidal perturbation}
\label{sec:diagonalization}

When $T\ped{tidal}\ll T\sped{lt}$, the cloud's structure is predominantly determined by the tidal potential of the companion, with the hyperfine splitting acting as a small perturbation. For this reason, it is now convenient to work in the basis of states $\ket{n\ell m'}$ aligned with the orbital angular momentum $\vec L_*$. From \eqref{eqn:Vbar-L*}, these states diagonalize the perturbations, with eigenvalues $\mathcal A-\mathcal C m'^2/T\ped{tidal}$. States with opposite $m'$ have thus the same energy. This degeneracy is broken by the small hyperfine interaction, with Hamiltonian
\beq
\mathcal H\sped{lt}=\frac{L_z}{T\sped{lt}}
=\frac{L_{z'}\cos\beta-L_{x'}\sin\beta}{T\sped{lt}}\,.
\label{eqn:hyperfine-primed}
\eeq
The $L_{x'}$ operator only connects states with $\Delta m'=\pm1$, so it always vanishes on the degenerate subspace. Only $L_{z'}$ is then left, under which the basis $\ket{n\ell m'}$ remains diagonal, with $\ket{n\ell m'}$ and $\ket{n\ell,\minus m'}$ acquiring an energy splitting of $2m'\cos\beta/T\sped{lt}$.\footnote{This splitting vanishes for $\cos\beta=0$. So, in a narrow band around $\beta=\pi/2$, the degeneracy must be broken by dissipative or higher order hyperfine interactions, and a different linear combination of $\ket{n\ell m'}$ and $\ket{n\ell,\minus m'}$ might be required to diagonalize the Hamiltonian. We set this sliver aside in the rest of the paper.} The diagonal part of the full Hamiltonian is then
\beq
\omega'_{n\ell m'}=\braket{n\ell m'|\mathcal H+\bar V|n\ell m'}=E_{n\ell0}+\mathcal A
-\frac{\mathcal C m'^2}{T\ped{tidal}}
+\frac{m'\cos\beta}{T\sped{lt}}
+i\Gamma'_{n\ell m'}\,,
\label{eqn:unlocked-frequencies}
\eeq
where
\beq
\Gamma'_{n\ell m'}=\sum_m\big[d^{(\ell)}_{mm'}(\beta)\big]^2\,\Gamma_{n\ell m}\,.
\label{eqn:unlocked-widths}
\eeq

\vskip 4pt
For the $(n,\ell)=(2,1)$ multiplet, the growth/decay rates of the orbit-aligned $\ket{n\ell m'}$ eigenstates are
\begin{align}
\Gamma'_{21\pm1}&=\bigg(\frac{1\pm\cos\beta}2\bigg)^2\Gamma_{211}+\frac{\sin^2\beta}2\Gamma_{210}+\bigg(\frac{1\mp\cos\beta}2\bigg)^2\Gamma_{21\minus1}\,,\\
\Gamma'_{210}&=\cos^2\beta\,\Gamma_{210}+\frac{\sin^2\beta}2(\Gamma_{211}+\Gamma_{21\minus1})\,,
\end{align}
while for the $(3,2)$ multiplet they are
\begin{align}
\Gamma'_{32\pm2}&=\bigg(\frac{1\pm\cos\beta}2\bigg)^4\Gamma_{322}+\bigg(\frac{1\pm\cos\beta}2\bigg)^2\sin^2\beta\,\Gamma_{321}+\frac38\sin^4\beta\,\Gamma_{320}\notag\\
&\quad+\bigg(\frac{1\mp\cos\beta}2\bigg)^2\sin^2\beta\,\Gamma_{32\minus1}+\bigg(\frac{1\mp\cos\beta}2\bigg)^4\Gamma_{32\minus2}\,,\\
\Gamma'_{32\pm1}&=\bigg(\frac{1\pm\cos\beta}2\bigg)^2\sin^2\beta\,\Gamma_{322}+\bigg(\frac{1\pm\cos\beta}2\bigg)^2(2\cos\beta\mp1)^2\Gamma_{321}+\frac38\sin^22\beta\,\Gamma_{320}\notag\\
&\quad+\bigg(\frac{1\mp\cos\beta}2\bigg)^2(2\cos\beta\pm1)^2\Gamma_{32\minus1}+\bigg(\frac{1\mp\cos\beta}2\bigg)^2\sin^2\beta\,\Gamma_{32\minus2}\,,\\
\Gamma'_{320}&=\frac38\sin^4\beta\,(\Gamma_{322}+\Gamma_{32\minus2})+\frac38\sin^22\beta\,(\Gamma_{321}+\Gamma_{32\minus1})+\bigg(\frac{3\cos^2\beta-1}2\bigg)^2\Gamma_{320}\,.
\end{align}
As in the more familiar BH-aligned basis $\ket{n\ell m}$, different eigenstates have different growth/decay rates, so after a time longer than $T\ped{abs}$ only the state with the largest $\Gamma'_{n\ell m'}$ (either positive, or least negative) survives. The coefficients $[d^{(\ell)}_{mm'}(\beta)]^2$ define a probability distribution over $m$ with mean $\braket{n\ell m'|L_z|n\ell m'}=m'\cos\beta$, so each $\Gamma'_{n\ell m'}$ is an average of the $\Gamma_{n\ell m}$ weighted towards $m\simeq m'\cos\beta$. As $\Gamma_{n\ell m}$ increases with $m$, the largest width is generally attained at $m'=\ell$ for $\beta<\pi/2$ and at $m'=-\ell$ for $\beta>\pi/2$, as one can check from the expressions above.

\vskip 4pt
We can thus make the simplifying assumption that the cloud is in a pure $\ket{n\ell m'}$ state with $m'=\ell$ (or $-\ell$ for $\beta>\pi/2$). In other words, the cloud is in the \emph{same} state that would be generated by superradiance, but aligned with the orbital angular momentum $\vec L_*$ rather than the BH spin $\vec S$. For this reason, we call this the \emph{unlocked} regime. Different from the familiar case of a spin-aligned cloud, however, in the unlocked regime the cloud can still decay ($\Gamma'_{n\ell m'}<0$) even if the BH's spin is high enough to satisfy the superradiant condition.

\subsection{Backreaction and evolution}
\label{sec:unlocked-bh-backreaction}

Assuming that the cloud is in a pure $\ket{n\ell m'}$ state, we can use the master equations from Sec.~\ref{sec:master-equations} to compute the evolution of the system. Combining the mass conservation \eqref{eqn:dotalpha-reduced} with \eqref{eqn:master-Mdot} gives
\beq
\dot X=2X(1+X)\sum_m\Gamma_{n\ell m}\big[d^{(\ell)}_{mm'}(\beta)\big]^2=2X(1+X)\Gamma'_{n\ell m'}\,,
\label{eqn:dotX-unlocked}
\eeq
while the evolution of the BH spin is found from \eqref{eqn:rotated-nlm} and \eqref{eqn:master-Sdot},
\beq
\dot S=-\frac{2XM^2}{\alpha}\sum_m m\,\Gamma_{n\ell m}\big[d^{(\ell)}_{mm'}(\beta)\big]^2\,.
\label{eqn:dotS-unlocked}
\eeq
Unlike the locked co-precession regime, the mass and spin evolution are not linked together. In particular, the nonexistence of a superradiant threshold means that the growth of an unlocked cloud must be followed by a decay \emph{on the same timescale}, rather than on a parametrically longer time as in the case of threshold drift. To see that the cloud cannot reach saturation, suppose at a certain time we have $\dot M=\dot M\ped{c}=0$, that is, $\Gamma'_{n\ell m'}=0$. Using Detweiler's approximation \eqref{eqn:Gamma_nlm}, we can write the growth/decay rates in the form $\Gamma_{n\ell m}=A_{n\ell m}(m\Omega_+-E_{n\ell m})$, and
\beq
\Gamma'_{n\ell m'}=\bigg(\sum_m\big[d_{mm'}^{(\ell)}(\beta)\big]^2A_{n\ell m}\bigg)(\bar m\Omega_+-E)\,,\qquad\bar m\equiv\frac{\sum_m m\big[d_{mm'}^{(\ell)}(\beta)\big]^2A_{n\ell m}}{\sum_m\big[d_{mm'}^{(\ell)}(\beta)\big]^2A_{n\ell m}}\,,
\eeq
where we replaced $E_{n\ell m}$ with a common $E$ because the hyperfine splittings are negligible in the unlocked regime. An unlocked cloud's mass is thus stationary when it satisfies the usual superradiance condition $\bar m\Omega_+=E$, with the \emph{effective} azimuthal number $\bar m$. Using this to eliminate $E$ in \eqref{eqn:dotS-unlocked}, we can write
\beq
\dot S\big|_{\dot M=0}=-\frac{2M\ped{c}\Omega_+}\mu\sum_m\,A_{n\ell m}\big[d_{mm'}^{(\ell)}(\beta)\big]^2(m-\bar m)^2\le0\,.
\label{eqn:dotS-on-dotM=0}
\eeq
This expression is negative definite and can vanish only when the sum includes a single nonzero term, with $m=\bar m$. Given the appearance of the Wigner matrix in \eqref{eqn:dotS-on-dotM=0}, this requires $\beta=0$ or $\beta=\pi$.\footnote{The same conclusion can be reached without using Detweiler's approximation. Indeed, the growth/decay of a single $\ket{n\ell m}$ mode saturates the area theorem inequality $\dot M-\Omega_+\dot S\ge0$ only on its superradiant threshold. An oblique orbit forces multiple modes to be active at the same time, but no two of them can be on the threshold at the same time. Hence $\dot M=0$ forces $\dot S<0$.} If the change in $\beta$ is neglected, a decreasing BH spin will then increase the decay rates (and decrease the growth rate of superradiant states), leading to cloud dissipation, $\dot M\ped{c}<0$ and $\dot M>0$.

\vskip 4pt
To determine the evolution of the obliquity, we start again from angular momentum conservation,
\beq
\vec J=\vec S+\bigg(L_*+\frac{m'XM^2}\alpha\bigg)\hat L_*\,,
\label{eqn:unlocked-J}
\eeq
where we used $\vec S\ped{c}=(m'XM^2/\alpha)\hat L_*$. Squaring this relation gives
\beq
\cos\beta=
\frac{J^2-S^2-(L_*+m'XM^2/\alpha)^2}{2S(L_*+m'XM^2/\alpha)}\,,
\label{eqn:unlocked-beta-X}
\eeq
where $J$ and $L_*$ are constants, while the BH mass $M$ and spin $S$ and the cloud mass fraction $X$ evolve according to \eqref{eqn:dotalpha-reduced}, \eqref{eqn:dotX-unlocked}, \eqref{eqn:dotS-unlocked}. Differentiating \eqref{eqn:unlocked-beta-X}, the rate of change of the obliquity is
\beq
\begin{split}
\dot\beta={}&\frac{1}{S(L_*+m'XM^2/\alpha)\sin\beta}\bigg[
\dot S\bigg(S+\bigg(L_*+\frac{m'XM^2}{\alpha}\bigg)\cos\beta\bigg)\\
&\qquad\qquad\qquad\qquad\qquad\qquad+\frac{m'M^2}{\alpha}\frac{\dot X}{1+X}
\bigg(L_*+\frac{m'XM^2}{\alpha}+S\cos\beta\bigg)\bigg]\,,
\end{split}
\label{eqn:dotbeta-unlocked}
\eeq
which in the limit of negligible orbital backreaction (large $L_*$) reduces to
\beq
\dot\beta=\frac{2XM^2}{\alpha S\sin\beta}
\sum_m(m'-m\cos\beta)\,
\Gamma_{n\ell m}\big[d^{(\ell)}_{mm'}(\beta)\big]^2\,.
\label{eqn:dotbeta-unlocked-reservoir}
\eeq

\vskip 4pt
To determine the azimuth evolution, we note that $S_{\text{c},x}=(m'XM^2/\alpha)\sin\beta$, $S_{\text{c},y}=0$, and
\beq
\tau_y=-\frac{iM\ped{c,0}}{\mu}\braket{c'|[\Re\mathcal H,L_y]|c'}=-\frac{iM\ped{c,0}}{\mu}\braket{c'|[\mathcal H\sped{lt},L_y]|c'}
=-\frac{S_{\text{c},x}}{T\sped{lt}}\,,
\eeq
where we used $[L_z,L_y]=-iL_x$ and $\braket{c'|\{\Im\mathcal H,L_y\}|c'}=0$. Plugging this in \eqref{eqn:master-varphi}, we obtain
\beq
\dot\varphi=\frac{J}{L_*+m'XM^2/\alpha}
\bigg(\gamma\ped{dS}L_*+\frac{m'XM^2}{\alpha S\,T\sped{lt}}\bigg)\,.
\label{eqn:dotvarphi-unlocked}
\eeq
Apart from the geodetic precession, we see that in the limit of negligible backreaction (large $L_*$) \eqref{eqn:dotvarphi-unlocked} reduces to $\dot\varphi=T\sped{lt}^{-1}S\ped{c}/S$. This result mirrors (up to constant prefactors) the one from locked co-precession, \eqref{eqn:dotphi-locked}, with the replacement $T\ped{tidal}\to T\sped{lt}$, as now it is the hyperfine coupling that drives the precession, rather than the tidal force.

\subsection{Final state: spin-orbit (anti-)alignment or cloud exhaustion}
\label{sec:unlocked-final}

Equation~\eqref{eqn:dotS-on-dotM=0} implies that a surviving cloud can reach a stationary state only when $\beta=0$ or $\pi$. Requiring $\ket{n\ell m'}$ to be the state with largest $\Gamma'_{n\ell m'}$, we must have $m'>0$ for $\beta=0$ and $m'<0$ for $\beta=\pi$. The other possible outcome is cloud exhaustion, just as in the locked case. As in Sec.~\ref{sec:final_locked}, we can use conservation laws to study either outcome. In this case, from \eqref{eqn:unlocked-J} the magnitude of the total angular momentum as a function of the initial parameters is
\beq
J=\sqrt{S_0^2+(L_*+\sgn(m')S\ped{c,0})^2+2S_0(L_*+\sgn(m')S\ped{c,0})\cos\beta_0}\,,
\label{eqn:J-unlocked}
\eeq
where $S_0=\tilde a_0M_0^2$ and $S\ped{c,0}=\abs{m'}X_0M_0^2/\alpha_0$. However, in the unlocked regime we cannot fix $S$ as a function of $M$ with the superradiant threshold condition throughout the entire evolution, so determining which outcome is reached generally requires integrating the reduced equations.

\paragraph{Spin (anti-)alignment.} If the system's final obliquity is $\beta_f=0$ or $\pi$ with a nonzero cloud mass fraction $X_f\ne0$, then the stationary final state requires the BH to live on the superradiant threshold \eqref{eqn:threshold}. Angular momentum conservation then gives the same expression for $X_f$ derived in the locked co-precession regime. The final cloud fraction $X_f$ is thus given by Eq.~\eqref{eqn:Xf-locked}, with $m_a$ replaced by $\abs{m'}$, and $\abs{J-L_*}$ replaced by
\beq
\begin{cases}
J-L_* & \text{for }\beta_f=0\text{ (i.e., }m'>0\text{)}\,,\\
L_*-J & \text{for }\beta_f=\pi\text{ (i.e., }m'<0\text{)}\text{ and }L_*>S_f+S_{\text{c},f}\,,\\
J+L_* & \text{for }\beta_f=\pi\text{ (i.e., }m'<0\text{)}\text{ and }L_*<S_f+S_{\text{c},f}\,.\\
\end{cases}
\label{eqn:branches}
\eeq

\vskip 4pt
In Sec.~\ref{sec:final_locked} we argued, by combining the triangle inequality with the fact that $S+S\ped{c}$ decreases, that the $J+L_*$ branch is not reachable. The same monotonicity holds here. At $\beta=\pi$ the orbit-aligned state $\ket{n\ell,\minus\ell'}$ coincides with the BH-aligned $\ket{a}$, and for a cloud close to it \eqref{eqn:master-Sdot} gives
\beq
\frac{\dd}{\dd t}\bigg(S+\frac{m_aM\ped{c}}\mu\bigg)=\frac{2M\ped{c,0}}\mu\sum_{b\ne a}(m_a-m_b)\Gamma_{b}\abs{c_b}^2\le0\,,
\label{eqn:SpSc-monotonic}
\eeq
because, on the threshold of $\ket{a}$, $\Gamma_b>0$ only if $m_b>m_a$. What changes is the geometry, as the cloud is now aligned with the orbit rather than with the BH spin. Expanding \eqref{eqn:unlocked-beta-X} at small $\pi-\beta$, with $m'<0$, gives
\beq
S+S\ped{c}-(J+L_*)\simeq\frac{S\,(S\ped{c}-L_*)}{2(S+S\ped{c}-L_*)}\,(\pi-\beta)^2\,.
\label{eqn:JpL-geometry}
\eeq
For $S\ped{c}<L_*<S+S\ped{c}$ the right-hand side is negative: $S+S\ped{c}$ lies below its endpoint value and, since it can only decrease, the endpoint cannot be reached, as in the locked regime. For $L_*<S\ped{c}$, instead, $S+S\ped{c}$ lies \emph{above} $J+L_*$, and the single-state evolution of Sec.~\ref{sec:unlocked-bh-backreaction} is attracted to the endpoint. In this case, however, the anti-aligned state is unstable to a mode that the single-state reduction does not contain: a tilt of the cloud relative to the orbit. The mechanism can be described as follows. Suppose the cloud is tilted by a small angle $\delta$ relative to the BH. Because the tidal interaction conserves $\vec S\ped{c}+\vec L_*$, the orbital axis reacts by tilting by an angle $(S\ped{c}/L_*)\delta$ relative to the BH. For $L_*<S\ped{c}$ the cloud thus lies between the BH spin and the orbital axis. Absorption then pulls the cloud towards the BH, and therefore \emph{away} from $-\hat L_*$, so the relative tilt grows. A simple linear analysis, which for brevity we do not report here, shows that the cloud-orbit tilt grows as $e^{t/T\ped{flip}}$, with
\beq
T\ped{flip}=\frac{S\ped{c}-L_*}{L_*}\,T\ped{abs}\,.
\label{eqn:T_flip}
\eeq
The $J+L_*$ branch of \eqref{eqn:branches} is therefore never realized, and we ignore it in what follows. The instability drives the system away from anti-alignment and towards alignment: we refer to this as an \emph{orbit flip}, and verify it by direct integration of the master equations in Sec.~\ref{sec:numerics}.

\vskip 4pt
Even though, for the first two branches in \eqref{eqn:branches}, the formula for the final cloud mass fraction $X_f$ in terms of $J$ and $L_*$ is identical to Eq.~\eqref{eqn:Xf-locked}, the blue and green contours in Fig.~\ref{fig:final-state} do not apply in the unlocked regime, because the expression of $J$ is now different, see \eqref{eqn:J-unlocked}.

\paragraph{Cloud exhaustion.} If the cloud is exhausted, we have $X_f=0$ and $M_f=M_0(1+X_0)$ from mass conservation. The final BH spin $S_f$, however, must be obtained by integrating the spin evolution equation \eqref{eqn:dotS-unlocked}. Once that is known, angular momentum conservation then fixes the final obliquity,
\beq
\cos\beta_f=\frac{J^2-S_f^2-L_*^2}{2S_fL_*}\,.
\label{eqn:unlocked-exhausted-endpoint}
\eeq

\begin{figure}
\centering
\includegraphics[width=\textwidth]{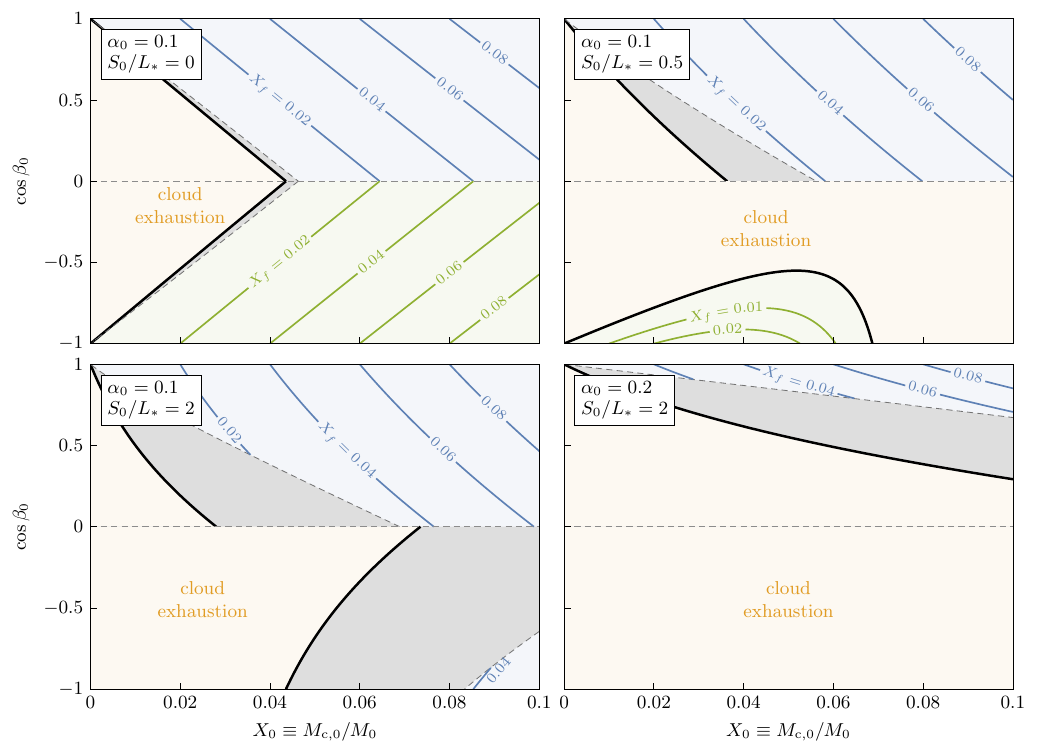}
\caption{Same as Fig.~\ref{fig:final-state}, but for the unlocked regime, based on Eqs.~\eqref{eqn:Xf-locked}, \eqref{eqn:J-unlocked}, and \eqref{eqn:branches}. All panels assume that the cloud starts on the superradiant threshold, with $\tilde a_0=4\alpha_0/(1+4\alpha_0^2)$, $(n,\ell)=(2,1)$, and $\abs{m'}=1$. More specifically, given the requirement of $\ket{n\ell m'}$ being the state with the largest $\Gamma'_{n\ell m'}$, we have $m'=1$ for $\cos\beta_0>0$ and $m'=-1$ for $\cos\beta_0<0$. We color with \textcolor{Mathematica1}{blue}, \textcolor{Mathematica3}{green}, and \textcolor{Mathematica2}{orange} backgrounds the regions where the final state stability arguments of the main text establish the outcome to be spin alignment, spin anti-alignment, or cloud exhaustion, respectively.  The gray areas mark instead the regions where those arguments alone cannot determine the final state. While the final obliquity in the case of cloud exhaustion cannot be predicted just from conservation laws (hence we do not draw \textcolor{Mathematica2}{orange} contours), the final cloud mass fraction $X_f$ can be calculated because (anti-)aligned systems must end up on the superradiant threshold, providing an additional constraint.}
\label{fig:final-state-unlocked}
\end{figure}

\vskip 4pt
We show in Fig.~\ref{fig:final-state-unlocked} the final state as a function of $X_0$ and $\beta_0$. We display four cases that span $\alpha_0=0.1$, $\alpha_0=0.2$, and different values of the ratio $S_0/L_*$, which quantifies the strength of the backreaction in the unlocked regime. As anticipated, to determine the actual final state of the system one in general has to integrate \eqref{eqn:dotX-unlocked}--\eqref{eqn:dotS-unlocked}. However, there are cases where we can determine the outcome by exclusion. For example, if the cloud mass fraction implied by both spin alignment and anti-alignment (the first two branches of \eqref{eqn:branches}) is $X_f\le0$, then the only possible final state is cloud exhaustion. We show that as a region with uniform orange background in Fig.~\ref{fig:final-state-unlocked}, without contours for $\cos\beta_f$ because expressing \eqref{eqn:unlocked-exhausted-endpoint} as a function of the initial parameters requires direct integration of the dynamics. Conversely, if one assumes $X_f=0$ but every possible BH final spin $S_f$ gives at least one orbit-aligned state with $\Gamma'_{n\ell m'}>0$, then cloud exhaustion is not a stable final state, and we must have spin (anti-)alignment. We show this alternative as a blue (green) region in Fig.~\ref{fig:final-state-unlocked}, where we constrain the possible final values of $S_f$ by requiring $\abs{\cos\beta_f}\le1$ and the BH area theorem inequality,
\beq
M_f^2+\sqrt{M_f^4-S_f^2}\ge M_0^2+\sqrt{M_0^4-S_0^2}\,,\qquad M_f=(1+X_0)M_0\,.
\eeq
When neither of the arguments above applies, the final state cannot be determined by these endpoint arguments, and a direct integration of the equations is needed, see Sec.~\ref{sec:numerics}. We color the corresponding areas with a gray background in Fig.~\ref{fig:final-state-unlocked}. 

\subsection{Alignment timescale}
\label{sec:unlocked-timescales}

The characteristic timescale for cloud depletion in the unlocked regime follows from \eqref{eqn:dotX-unlocked}. As in Sec.~\ref{sec:locked-timescales}, close to alignment this reads $\dot X\simeq-X\sin^2\beta/T\ped{depl}$, with
\beq
T\ped{depl}=\frac{T\ped{abs}}{\abs{m'}}\,.
\label{eqn:Tdepl-unlocked}
\eeq
This result is easily interpretable: an unlocked cloud consists of a dominant state, generically living away from the superradiant threshold, hence it decays over the bare absorption timescale $T\ped{abs}$, as opposed to the suppressed rate \eqref{eqn:T_align} found in the locked co-precession regime.

\vskip 4pt
The alignment timescale $T\ped{align}$ comes instead from \eqref{eqn:dotbeta-unlocked}. Close to alignment this gives $\dot\beta\simeq-\sin\beta/T\ped{align}$, with
\beq
T\ped{align}=\bigg(\frac{S\ped{c}}S+\frac{\sgn(m')S\ped{c}}{L_*+\sgn(m')S\ped{c}}\bigg)^{-1}T\ped{abs}\,.
\label{eqn:Talign-unlocked}
\eeq
This goes from $T\ped{align}\simeq(S/S\ped{c})T\ped{abs}$ for small backreaction to $T\ped{align}\simeq T\ped{abs}/(1+S\ped{c}/S)$ for large backreaction. The weak dependence on the backreaction is due to the fact that in the unlocked regime the orbit and the cloud are already aligned, hence spin (anti-)alignment simply requires the BH's spin to rotate. Note that, for $m'<0$, $T\ped{align}$ is negative for $S\ped{c}<L_*<S+S\ped{c}$, meaning that the system moves away from anti-alignment.

\vskip 4pt
Finally, the characteristic precession timescale follows from \eqref{eqn:dotvarphi-unlocked},
\beq
T\ped{prec}=\frac{T\ped{align}}{T\ped{abs}}\,T\sped{lt}\,,
\label{eqn:Tprec-unlocked}
\eeq
where we neglected geodetic precession. As anticipated, this result has a simple physical interpretation, namely that the Lense--Thirring Hamiltonian dominates the cloud-BH conservative interaction in the unlocked regime.

\section{Numerical integration}
\label{sec:numerics}

The analytical results of Secs.~\ref{sec:locked} and~\ref{sec:unlocked} rely on approximations such as adiabatic elimination and threshold drift in the locked regime, and the restriction to a single orbit-aligned state in the unlocked one. We now test these approximations against a direct numerical integration of the master equations. We describe the setup in Sec.~\ref{sec:numerics-implementation}, and compare the results with the reduced equations and the predicted final states in Secs.~\ref{sec:numerics-locked} and~\ref{sec:numerics-unlocked}, for the locked and unlocked regimes, respectively. In the latter, we also verify that the $J+L_*$ branch is unstable, as argued in Sec.~\ref{sec:unlocked-final}.

\subsection{Setup}
\label{sec:numerics-implementation}

We solve numerically the master equations \eqref{eqn:master-ca}--\eqref{eqn:master-varphi} retaining all five hydrogenic states with $n\le2$, and use the matrix elements of the quadrupole \eqref{eqn:Vbar-quad}. Including states with higher $n$ is possible, but it massively slows down the integration due to the need to resolve the short Bohr timescale. We checked, in a few trial integrations, that the inclusion of higher-$n$ states does not appreciably change the results; hence we prefer to stick with $n\le2$ for the integrations reported here. To calculate the eigenenergies $\omega_{n\ell m}$ we use Leaver's method~\cite{Leaver:1985ax,Dolan:2007mj,Baumann:2019eav}: this is especially important at moderate-to-large $\alpha$, where the analytical formula \eqref{eqn:Gamma_nlm} introduces conspicuous errors. We perform two sets of numerical integrations, one for the locked co-precession regime, and one for the unlocked regime.

\begin{table}
\centering
\begin{tabular}{lcccccccc}
\toprule
& $\alpha_0$ & $X_0$ & $\cos\beta_0$ &  & $\dfrac{T\sped{lt,0}}{T\ped{tidal,0}}$ & $q$ & $\frac{a}{M_0}$ & final state \\
\midrule
\multicolumn{4}{l}{\textbf{Locked}} & $\mathcal B_0$ & & & & \\
alignment & $0.2$ & $0.08$ & $0.85$ & $0.04$ & $0.02$ & $1.01$ & $1148$ & $X_f\simeq 0.024$ \\
exhaustion & $0.2$ & $0.08$ & $0.65$ & $0.04$ & $0.02$ & $1.01$ & $1148$ & $\cos\beta_f\simeq 0.83$ \\
anti-alignment & $0.2$ & $0.1$ & $-0.92$ & $0.04$ & $0.02$ & $1.08$ & $1180$ & $X_f\simeq 0.063$ \\
\midrule
\multicolumn{4}{l}{\textbf{Unlocked}} & $S\ped{c,0}/L_*$ & & & & \\
alignment & $0.08$ & $0.0128$ & $0.75$ & $4.2$ & $5$ & $0.0017$ & $460$ & $X_f\simeq 0.010$ \\
exhaustion & $0.08$ & $0.0018$ & $-0.9$ & $0.60$ & $5$ & $0.0017$ & $460$ & $\cos\beta_f\simeq -0.28$ \\
orbit flip \#1 & $0.08$ & $0.01$ & $-0.9$ & $3.3$ & $5$ & $0.0017$ & $460$ & $X_f\simeq 0.0029$ \\
orbit flip \#2 & $0.08$ & $0.006$ & $-0.6$ & $2.0$ & $5$ & $0.0017$ & $460$ & $\cos\beta_f\simeq1$ \\
\bottomrule
\end{tabular}
\caption{Initial parameters and final states of the numerical integrations.}
\label{tab:runs}
\end{table}

\vskip 4pt
In the locked case, we initialize the cloud in the $\ket{211}$ state, with the initial amplitudes of the other states set equal to the adiabatic elimination formula \eqref{eqn:adiabatic-elimination}---this choice removes an initial transient where subdominant states would get excited and then relax to \eqref{eqn:adiabatic-elimination}. We pick three different initial conditions, corresponding to the three possible final states: spin alignment, cloud exhaustion, and spin anti-alignment. All of them have $\alpha_0=0.2$, $\mathcal B_0=0.04$, and $T\sped{lt,0}/T\ped{tidal,0}=0.02$, where we denote with a subscript ``0'' the characteristic timescales evaluated at the initial state. From those parameters, $q$ and $a$ can be derived from \eqref{eqn:orbital-definitions} and \eqref{eqn:T_tidal} assuming $e=0$. In the unlocked case, we start from the orbit-aligned eigenstate $\ket{21m'_0}$, where $m'_0=\sgn(\cos\beta_0)$. Here we fix $\alpha_0=0.08$, $q\simeq0.0017$ and $T\ped{tidal,0}/T\sped{lt,0}=0.2$ (giving $S_0/L_*\simeq8.3$), from which $a$ can be derived, and choose four different initial conditions. Table~\ref{tab:runs} summarizes the initial parameters.

\vskip 4pt
In all cases the initial BH spin is on the $\ket{211}$ superradiant threshold, such that $\Gamma_{211}=0$. From the chosen initial conditions, parity selection rules keep the states $\ket{100}$ and $\ket{200}$ always empty, so only the $\ket{21m}$ triplet is populated. Also, we note that the semi-major axis corresponding to the chosen parameters is $a\simeq46r\ped{c,0}$ in the locked cases and $a\simeq3r\ped{c,0}$ in the unlocked cases. The quadrupolar approximation \eqref{eqn:Vbar-quad} for the companion's potential is thus expected to be accurate for the locked runs, but not necessarily for the unlocked ones. Furthermore, a companion inside the cloud would trigger important dynamical effects such as ionization \cite{Baumann:2021fkf} and resonances \cite{Baumann:2019ztm}. We prefer nevertheless to keep these parameters as they combine ease of numerical integration with insightful results.

\vskip 4pt
We also numerically integrate the reduced systems of equations from Secs.~\ref{sec:reduced-system} and~\ref{sec:unlocked-bh-backreaction}, and show the results alongside the answer from the full master equations. To make a proper comparison, we use Leaver's method to evaluate the decay rates $\Gamma_b$ appearing in the reduced equations, as well as to determine the position of the superradiant threshold. However, we keep for simplicity the analytical form of the prefactor \eqref{eqn:kappa} appearing in the reduced equations of locked co-precession, which was obtained from the analytical approximation of the threshold \eqref{eqn:threshold}.

\subsection{Locked co-precession}
\label{sec:numerics-locked}

\begin{figure}
\centering
\includegraphics[width=\textwidth]{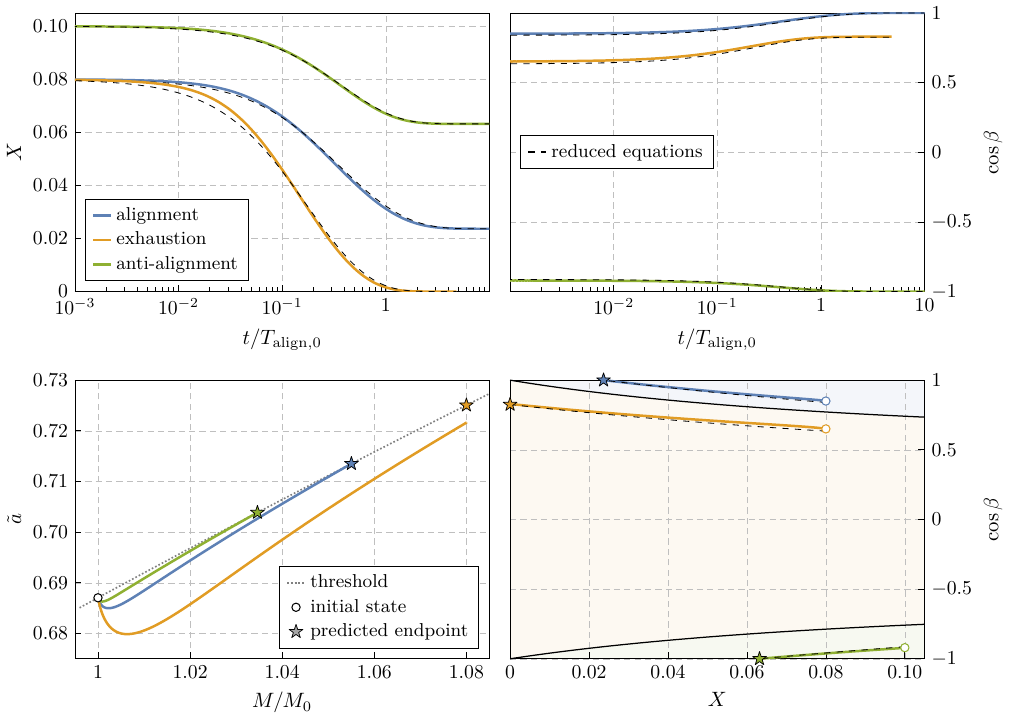}
\caption{Direct integration of the master equations in the locked co-precession regime, for the three sets of initial parameters from Table~\ref{tab:runs}. The top panels show the time evolution of the cloud mass fraction $X$ and BH-orbit obliquity $\beta$. In the bottom panels we show the evolution of the system in the $(M/M_0,\tilde a)$ and $(X,\cos\beta)$ planes. In all panels, dashed black lines denote the evolution predicted by the reduced system of equations derived in Sec.~\ref{sec:reduced-system}. The colored stars are the final states predicted in Sec.~\ref{sec:final_locked}, while the background colors in the bottom right panel denote the predicted regions of spin (anti-)alignment and cloud exhaustion, with the same color code as in Fig.~\ref{fig:final-state}.}
\label{fig:numerics-locked}
\end{figure}

We show in Fig.~\ref{fig:numerics-locked} the results of the direct integration in the locked cases. In all three cases, the solution tracks reliably the prediction from the one-dimensional reduced system of equations. On a timescale of order $\mathcal O(T\ped{align,0})$, the systems reach the endpoints predicted in Sec.~\ref{sec:final_locked} with good accuracy. The only meaningful deviation from the analytical treatment of Sec.~\ref{sec:reduced-system} is a gap between the superradiant threshold and the actual BH trajectory, which develops at the beginning of the integration and persists throughout. This is expected, because the cloud's dissipation requires $\Gamma_{211}<0$, which is only possible below the threshold. The gap is of order $\tilde a\sim\mathcal O(0.01)$ for the run ending with cloud exhaustion, in which case it persists until the final state is reached; it is smaller for the other two cases, in both of which the final state necessarily sits on the threshold because the cloud survives.

\subsection{Unlocked alignment}
\label{sec:numerics-unlocked}

\begin{figure}
\centering
\includegraphics[width=\textwidth]{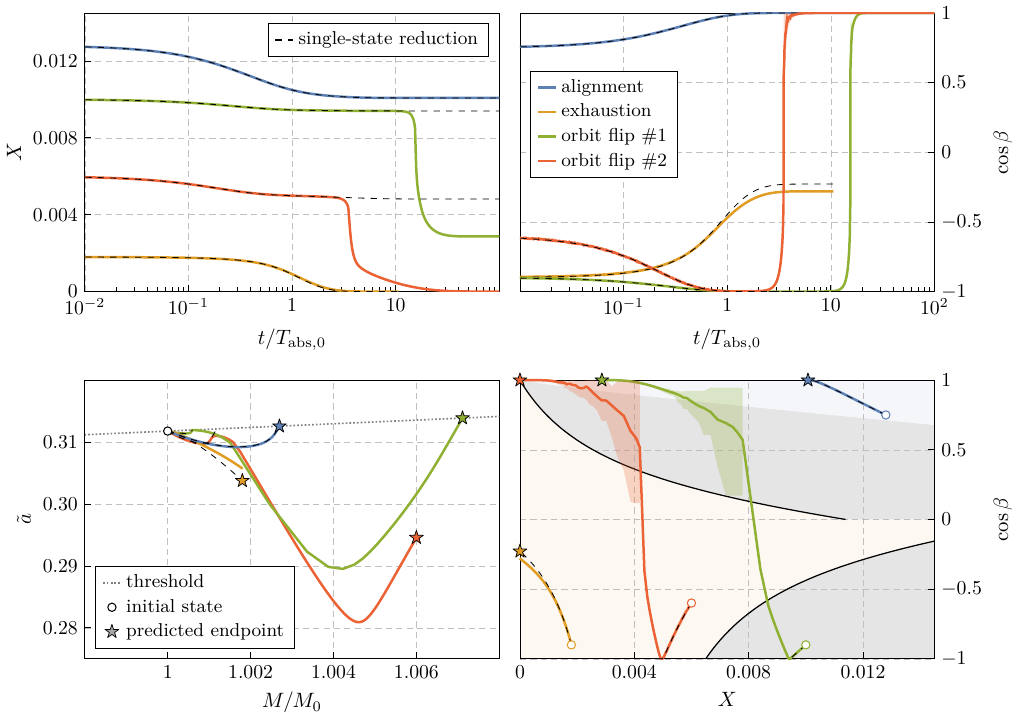}
\caption{Same as Fig.~\ref{fig:numerics-locked}, but for the four sets of parameters in the unlocked regime from Table~\ref{tab:runs}. The position of the \textcolor{Mathematica3}{green} star is computed conditional on the system reaching spin alignment (the system starts in the gray zone, where the analytical machinery of Sec.~\ref{sec:unlocked} cannot predict which final state is reached). The colored bands around the trajectories of the two orbit flip runs in the right panels denote the range of the rapid oscillations of the obliquity (nutation).}
\label{fig:numerics-unlocked}
\end{figure}

Figure~\ref{fig:numerics-unlocked} shows the results of the direct integration in the unlocked cases. For the first two sets of initial parameters in Table~\ref{tab:runs}, the numerical solutions closely follow the predictions of Sec.~\ref{sec:unlocked-bh-backreaction} and reach spin alignment or cloud exhaustion on a timescale of order $\mathcal O(T\ped{abs,0})$. We emphasize however that here the ``reduced system'' is not one-dimensional as in the locked case, and simply amounts to the single state $\ket{21m_0'}$ approximation. Because in this case we do not initialize the other states with a nonzero amplitude, there is an initial transient (not visible in Fig.~\ref{fig:numerics-unlocked}) where the cloud relaxes to a state that includes the perturbation from the hyperfine interaction.

\vskip 4pt
The outcome of the two other runs, which we label as ``orbit flip'', is richer. As Fig.~\ref{fig:numerics-unlocked} demonstrates, these cases initially approach anti-alignment ($\cos\beta=-1$) while accurately tracking the evolution predicted by Sec.~\ref{sec:unlocked-bh-backreaction}. However, they subsequently develop an instability which induces $\cos\beta$ to change sign and begin rapid and violent oscillations, which can be interpreted as spin-orbit nutation. This is the instability of the $J+L_*$ branch of \eqref{eqn:branches} discussed in Sec.~\ref{sec:unlocked-final}: both runs have $S\ped{c,0}>L_*$, so the system is first attracted to anti-alignment, until the cloud--orbit relative tilt grows large enough to flip the orbit. The system then quickly reaches spin alignment ($\cos\beta=1$) and the cloud decays, either partially ($X_f>0$, in the first orbit flip run) or completely ($X_f=0$, in the second orbit flip run).

\vskip 4pt
It is also interesting to note that the exhaustion run has the obliquity evolving away from anti-alignment, opposite to what is seen in the two orbit flip runs, which also start with $\cos\beta_0<0$. The reason is that, for the exhaustion run, we have $S\ped{c,0}<L_*<S+S\ped{c}$, giving a negative $T\ped{align}$ from \eqref{eqn:Talign-unlocked}, meaning that the system must move away from anti-alignment. The two orbit flip runs have instead $S\ped{c}>L_*$, so $T\ped{align}$ is positive and they initially move towards anti-alignment.

\section{Timescale competition: alignment vs gravitational-wave inspiral}
\label{sec:alignment-merger}

We have so far studied the dynamics of spin (anti-)alignment and cloud exhaustion ignoring other physical processes that may occur at the same time. The characteristic timescales derived in Secs.~\ref{sec:locked-timescales} and~\ref{sec:unlocked-timescales} should thus be compared with other dynamical timescales. We consider here a particularly relevant one, namely the merger time due to GW emission, which for $e=0$ is\footnote{All comparisons of timescales should be intended up to $\mathcal O(1)$ factors. For example, the instantaneous timescale of orbital decay, $a/\dot a$, is larger than $T\ped{merger}$ by a factor of 4. Using $a/\dot a$ in place of $T\ped{merger}$ would thus slightly enlarge the region where the spin alignment dynamics dominates over GW decay.}~\cite{Peters:1964zz}
\beq
T\ped{merger}=\frac{5a^4}{256\,q(1+q)(M+M\ped{c})^3}\,.
\label{eqn:T_merger}
\eeq

\begin{figure}
\centering
\includegraphics[width=\textwidth]{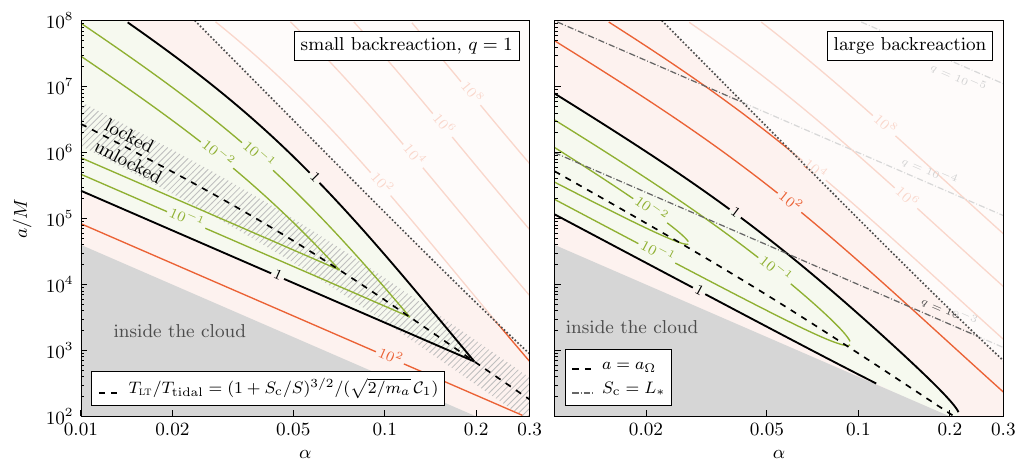}
\caption{Ratio $T\ped{align}/T\ped{merger}$ for $\ket{211}$ and $X=0.01$, in the regimes of small (left, assuming $q=1$) and large (right) orbital backreaction. In the left panel, the dashed line corresponds to $T\sped{lt}/T\ped{tidal}\simeq(1+S\ped{c}/S)^{3/2}/(\sqrt{2/m_a}\,\mathcal C_1)$; we use \eqref{eqn:ratio-locked-merger} above it, and \eqref{eqn:ratio-unlocked-merger} below it. The hatched band, which covers a decade in $T\sped{lt}/T\ped{tidal}$ on either side of the line, denotes the region where the analytical approximations for $T\ped{align}$ may be inaccurate. In the right panel, $T\ped{align}/T\ped{merger}$ is given by \eqref{eqn:ratio-strong-merger} and the dashed line corresponds to $a=a_\Omega$; the result is independent of $q$ as long as $S\ped{c}\gg L_*$, that is, well below the dash-dotted line corresponding to the chosen value of $q$. In both panels, the dotted line corresponds to the hyperfine resonance $\Omega=E_{211}-E_{210}$, given in \eqref{eqn:a-hyperfine}, evaluated for $q=1$ (left) and $q\ll1$ (right). Our results only hold below it, where the orbit average of Sec.~\ref{sec:tidal-perturbation} is valid, so we fade the colors above it. We color with \textcolor{Mathematica3}{green} (\textcolor{Mathematica4}{red}) backgrounds the regions where the ratio $T\ped{align}/T\ped{merger}$ is smaller (larger) than unity. Our results are only applicable in the \textcolor{Mathematica3}{green} region, as we neglect the orbital decay induced by GWs. In the gray region the companion is inside the cloud, $a<n^2M/\alpha^2$.}
\label{fig:alignment-merger}
\end{figure}

\vskip 4pt
Let us compare $T\ped{merger}$ with $T\ped{align}$, starting from the case of negligible orbital backreaction. In this section we focus on the $\ket{211}$ state only---as we pointed out in Secs.~\ref{sec:tidal-perturbation} and \ref{sec:drift}, for $\ket{a}=\ket{322}$ our treatment is expected to fail due to neglecting the coupling to $\ket{300}$. In the locked regime, we substitute \eqref{eqn:T_tidal}, \eqref{eqn:T_lt}, and \eqref{eqn:T-abs-explicit} into \eqref{eqn:T_align-small} to get
\beq
\frac{T\ped{align}}{T\ped{merger}}\bigg|\ped{locked}
\simeq\frac{1024}{1215}\frac{1+q}{q}\bigg(1+\frac{S\ped{c}}S\bigg)^2\bigg(1+\frac S{S\ped{c}}\bigg)\bigg(\frac aM\bigg)^2\frac{\alpha^{10}}{1+4\alpha^2}\,,
\label{eqn:ratio-locked-merger}
\eeq
while in the unlocked regime \eqref{eqn:Talign-unlocked} gives
\beq
\frac{T\ped{align}}{T\ped{merger}}\bigg|\ped{unlocked}
\simeq\frac{3072}5q(1+q)\frac{S}{S\ped{c}}\bigg(\frac Ma\bigg)^4\frac{1+4\alpha^2}{\alpha^{10}}\,.
\label{eqn:ratio-unlocked-merger}
\eeq
We show these ratios in the left panel of Fig.~\ref{fig:alignment-merger} for $X=0.01$. Note that in \eqref{eqn:ratio-locked-merger} and \eqref{eqn:ratio-unlocked-merger} we used here Detweiler's approximation to evaluate $T\ped{abs}$. This underestimates the decay rate $\Gamma_{n\ell m}$, meaning that we are deriving here a conservative set of conditions under which alignment acts faster than GW orbital decay. At fixed $\alpha$, \eqref{eqn:ratio-locked-merger} scales as $a^2$, while \eqref{eqn:ratio-unlocked-merger} scales as $a^{-4}$. As the semi-major axis is reduced by the GW emission, $T\ped{align}/T\ped{merger}$ thus decreases until the cloud unlocks, and increases afterwards.\footnote{Note that Fig.~\ref{fig:alignment-merger} is obtained using the formulae for the timescales derived at fixed $a$. We do not solve for the evolution with a shrinking semi-major axis, so we cannot study the transition from the locked to the unlocked regime during an actual inspiral. In particular, because $\ket{n\ell\ell}$ is the most energetic state in a BH-aligned multiplet, one can expect it to transition into a $\ket{n\ell0'}$ unlocked state (rather than $\ket{n\ell\ell'}$), as that is the most energetic one in the unlocked regime, see \eqref{eqn:unlocked-frequencies}.} The most favorable stage for alignment is therefore the transition between the two regimes, where the two ratios match, at $T\sped{lt}/T\ped{tidal}\simeq(1+S\ped{c}/S)^{3/2}/(\sqrt{2/m_a}\,\mathcal C_1)$. Evaluating the ratio there and expanding at small $\alpha$ gives
\beq
\frac{T\ped{align}}{T\ped{merger}}\bigg|\ped{min}
\simeq0.035\,\frac{1+q}{q^{1/3}}\bigg(1+\frac{S\ped{c}}S\bigg)\bigg(1+\frac S{S\ped{c}}\bigg)\bigg(\frac\alpha{0.2}\bigg)^{10/3}\,.
\label{eqn:ratio-crossover-merger}
\eeq
For $q\sim1$, we thus have $T\ped{align}\ll T\ped{merger}$ at some point in the inspiral for any $\alpha\lesssim0.2$.

\vskip 4pt
The results change when the orbital backreaction is large, which is typical for small $q$. From \eqref{eqn:T_align-strong}, we have
\beq
\frac{T\ped{align}}{T\ped{merger}}\simeq\frac{256X}{5\alpha}\,\frac{T\ped{abs}}M\bigg(\frac Ma\bigg)^{9/2}\bigg[1+\bigg(\frac a{a_\Omega}\bigg)^{7/2}\bigg]^2\,,
\label{eqn:ratio-strong-merger}
\eeq
which we show in the right panel of Fig.~\ref{fig:alignment-merger}. Remarkably, the mass ratio has dropped out. As the binary inspirals, the ratio \eqref{eqn:ratio-strong-merger} decreases as $a^{5/2}$ and then increases as $a^{-9/2}$, so the most favorable stage is now $a\sim a_\Omega$, rather than $T\sped{lt}\sim T\ped{tidal}$, with a minimum
\beq
\frac{T\ped{align}}{T\ped{merger}}\bigg|\ped{min}\simeq0.40\,\bigg(1+\frac{S\ped{c}}S\bigg)^{9/7}\bigg(\frac{X}{0.1}\bigg)^{-2/7}\bigg(\frac\alpha{0.2}\bigg)^{22/7}\,.
\label{eqn:ratio-strong-min}
\eeq
Similar to \eqref{eqn:ratio-crossover-merger}, alignment occurs at some point in the inspiral for $\alpha\lesssim0.2$. We note however that for small $q$, which in astrophysically relevant situations implies large $M$, the absolute timescales can become very large: for example, at the minimum in \eqref{eqn:ratio-strong-min}, with $\alpha=0.05$, $M=10^5M_\odot$, $X=0.01$, and $q=10^{-5}$ we get $T\ped{align}\sim\SI{3e9}{yr}$ and $T\ped{merger}\sim\SI{e11}{yr}$, and all timescales increase linearly with $M$.

\section{Conclusions}
\label{sec:conclusions}

Besides driving transitions between different states of the superradiant cloud at discrete resonant orbits, the tidal field of a binary companion also affects the gravitational atom through its static part. In this paper, we study the resulting evolution, treating the BH spin, the cloud, and the orbit as independent, mutually coupled vectors. On inclined orbits, the tide mixes the superradiant state with decaying ones: their horizon absorption depletes the cloud and, by angular momentum conservation, changes the BH obliquity relative to the binary plane. At large enough separations, the Lense--Thirring interaction locks the cloud to the BH spin, the BH undergoes threshold drift, and the dynamics reduces to a single, one-dimensional evolution equation. At smaller separations, the cloud aligns with the orbital axis rather than the BH spin, and depletes on the horizon absorption timescale.

\vskip 4pt
We find that, in both regimes, the system either reaches spin-orbit (anti-)alignment before the cloud is exhausted, or loses the cloud first, generally at an intermediate obliquity. In the locked regime, the outcome follows from conservation laws and the threshold condition: the spin \mbox{(anti-)aligns} if it starts close enough to the orbital axis or its opposite, or aligns from any obliquity under strong orbital backreaction. The phenomenology is a bit richer in the unlocked regime, with the possibility of an ``orbit flip'' from anti-alignment to alignment, under certain conditions. Numerical integrations of the master equations for a $\ket{211}$ cloud confirm the reduced dynamics we derived analytically in the two regimes, as well as the instability that leads to the orbit flip. Finally, we showed that alignment is faster than the GW-driven inspiral at some stage for a $\ket{211}$ cloud with $\alpha\lesssim0.2$.

\vskip 4pt
Our results rely on a number of simplifying assumptions. We considered only the orbit-averaged quadrupole of a companion outside the cloud and restricted the cloud to a single multiplet, which for $\ket{322}$ neglects the mixing with the rapidly decaying $\ket{300}$ state that dominates the depletion found in \cite{Fan:2023jjj} and could change our conclusions for this state. We also neglected three effects that should be included in a more complete treatment, namely the cloud's self-gravity \cite{Kim:2025wwj,Foster:2026nuz,Boskovic:2026boh}, quartic self-interactions \cite{Baryakhtar:2020gao,Takahashi:2024fyq,Witte:2024drg,
Cannizzaro:2026qgf}, and decay through GW emission \cite{Yoshino:2013ofa,Brito:2017zvb}. Collectively, we expect these effects to restrict to small $\alpha$ the applicability of our findings in old stellar-mass binaries. Also, in our setup and numerical integrations we kept the semi-major axis fixed, rather than following its GW-driven inspiral, so the possible transition between the locked and unlocked regimes, during which the cloud may also change state (see Sec.~\ref{sec:alignment-merger}), remains to be studied.

\vskip 4pt
The initial state we assumed, a BH on the superradiant threshold with a saturated cloud, is the natural one for a stellar-mass BH that pairs dynamically in a star cluster after its cloud has grown, especially at larger $\alpha$, where growth is faster. A BH formed in a supernova within an existing binary instead grows its cloud with the companion already present, a case we did not model, in which the companion could even prevent the growth altogether \cite{Tong:2022bbl,Zhu:2024bqs}.

\vskip 4pt
Several studies have computed the dynamics of gravitational atoms on inclined orbits deep in the inspiral \cite{Tomaselli:2023ysb,Cao:2024wby,Tomaselli:2025jfo,Boskovic:2025ixx}, recently also with a fully relativistic formalism \cite{Xu:2026aic,Xu:2026qrg}. Our results indicate that this is not what one should expect (unless the binary is formed below the green region in Fig.~\ref{fig:alignment-merger}): if resonances are neglected, a comparable-mass binary with a massive enough $\ket{211}$ cloud and $\alpha\lesssim0.2$ should reach the late inspiral either on an equatorial orbit, prograde or retrograde, with a surviving cloud, or without a cloud. At small mass ratios, orbital backreaction and the orbit flip favor prograde orbits, as also found in \cite{Fan:2023jjj} for a $\ket{322}$ cloud at fixed BH spin, although the timescales can then become very long.

\vskip 4pt
These findings complement those from \cite{Tomaselli:2024dbw,Boskovic:2025ixx} on the way resonances shape the possible final states of the system. Taken together, these phenomena raise the possibility of inferring the existence of clouds from a population of inspiralling binaries. Even for isotropic initial orientations, the mechanism studied in this paper could produce an abundance of (anti-)aligned spins in the tilt distribution constrained by the LIGO-Virgo-KAGRA collaboration \cite{KAGRA:2021duu,LIGOScientific:2025pvj}. In the locked regime, the BH spin magnitude also ends close to the superradiant threshold, a long-proposed signature of superradiance \cite{Arvanitaki:2010sy,Brito:2017zvb,Ng:2019jsx,Ning:2026ebu}, whether or not the cloud survives. One could then look for correlations between spin tilts, spin magnitudes, mass ratios, and BH masses, which set $\alpha$ for a given boson. Since other processes also align spins, dedicated population studies will be needed to turn this into a test of fundamental physics, as for the resonant imprints on eccentricity and obliquity \cite{Boskovic:2024fga,Tomaselli:2024dbw,Tomaselli:2024bdd,Boskovic:2025ixx}.

\vskip 4pt
Another interesting setup is one with multiple perturbers, such as a nuclear star cluster around a supermassive BH. It is natural to ask whether, in that case, all orbits align on the equatorial plane, and whether the observed distribution of inclinations \cite{Gillessen:2017jxc} can be used to constrain the existence of boson clouds, complementary to bounds from apsidal precession \cite{GRAVITY:2023cjt,Tomaselli:2025zdo}. Unfortunately, however, at the typical distance where the S-stars around Sgr~A* are located, the alignment timescale is far longer than the age of the Universe. Nevertheless, we consider the multi-perturber problem still worth investigating in future works, due to its potential relevance for other systems where the orbits are tighter and the alignment timescale shorter.

\vskip 4pt
As anticipated, the static tidal interaction is only one side of the problem. Since the orbit average requires an orbital frequency larger than the hyperfine splitting, the secular evolution studied here takes place at smaller separations than the hyperfine resonances, which the cloud must therefore survive, unless the binary forms at smaller separations. Survival is possible on nearly counter-rotating orbits \cite{Tomaselli:2024bdd,Tomaselli:2024dbw}, though not always for comparable masses \cite{Boskovic:2025ixx}. Such clouds could then be driven to exact anti-alignment or, with strong orbital backreaction, to alignment. A complete picture of the distribution of spin magnitudes and orientations of gravitational atoms in binaries will require simultaneously following resonances, ionization, and the evolution driven by the static tide all together, from binary formation to merger.

\paragraph{Acknowledgements.} G.M.T.\ gratefully acknowledges support from the Rubicon Fellowship, awarded by the Netherlands Organisation for Scientific Research (NWO), grant ID \url{https://doi.org/10.61686/WYKDB06497}. R.R.R.\ acknowledges financial support through the Science and Technology Facilities Council (STFC) grant ST/T00049X/1 and the IAS.

\paragraph{AI use.} We used Claude Code (Anthropic) and ChatGPT (OpenAI) during most stages of this work. An AI-generated GitHub repository that reproduces all analytical and numerical results of this paper can be found at \cite{spin-orbit-alignment-grav-atom-binaries:github}.

\clearpage
\phantomsection

\addtocontents{toc}{\protect\vskip24pt}
\addcontentsline{toc}{section}{References}

\makeatletter

\interlinepenalty=10000

{\linespread{1.0542}
\bibliographystyle{utphys}
\bibliography{main}
}
\makeatother

\end{document}